\documentclass[10pt,letterpaper]{article}
\usepackage[english]{babel}
\usepackage{graphicx}

\newcommand{\alt}{\mathbin{\lower 3pt\hbox
   {$\rlap{\raise 5pt\hbox{$\char'074$}}\mathchar"7218$}}}
\newcommand{\agt}{\mathbin{\lower 3pt\hbox
   {$\rlap{\raise 5pt\hbox{$\char'076$}}\mathchar"7218$}}}

\begin{document}
\setcounter{footnote}{0}
\setcounter{equation}{0}
\setcounter{figure}{0}
\setcounter{table}{0}
\vspace*{5mm}

\begin{center}
{\large\bf $T_c$ of disordered superconductors \\
near the  Anderson transition}

\vspace{4mm}
\vspace{4mm}
I. M. Suslov \\
Kapitza Institute for Physical Problems,
\\  Moscow, Russia \\
\vspace{1mm}
\end{center}

\begin{center}
\begin{minipage}{135mm}
{\small According to the Anderson theorem, the critical
temperature $T_c$ of a disordered superconductor
is determined by the average density of states
and does not change at the localization
threshold. This statement is valid  under assumption of a
self-averaging order parameter, which can be violated in the
strong localization region. Stimulating by statements on the
essential increase of $T_c$ near the Anderson transition, we
carried out the systematic investigation of  possible violations
of self-averaging. Strong deviations from the Anderson
theorem are possible due to resonances at the quasi-discrete
levels, resulting in localization of the order parameter at the
atomic scale. This effect is determined by the properties of
individual impurities and  has no direct relation to the Anderson
transition. In particular, we  do not see any reasons to say on
"fractal superconductivity" near the localization threshold.}
\end{minipage}
\end{center}

\twocolumn


\begin{center}
{\bf 1. Introduction}
\end{center}

The general picture of coexistence of superconductivity and
the Anderson localization was formed in the papers by
Bulaevskii and Sadovskii  \cite{1}--\cite{5} (see also
\cite{6,7}). According to the Anderson theorem \cite{8}, the
critical temperature $T_c$ of a disordered superconductor is
determined by the average density of states and does not
depend on the form of one-particle eigenstates. Since the
average density of states does not have singularity at the
Anderson transition, so  $T_c$ has the analogous behavior.
The coefficient of the gradient
term in the Ginzburg--Landau expansion, determining the
superconducting response of the system, remains finite at
the critical point. In the localized phase, the system breaks up
into quasi-independent blocks of  size  $\xi$ ($\xi$ is
the localization length) and superconductivity is suppressed due
to the size effect, when the average level spacing in such a
block becomes greater than  $T_c$.

Recently it was stated by Feigelman et al \cite{9,10} that  $T_c$
increases at approaching  the Anderson transition from the
metallic side and continues to grow in the localized phase
(with a maximum in the deep of it);
it is related with
multifractality of  wave functions. More
than that, $T_c$ depends on the Cooper interaction constant  $g$
not exponentially, but in the power-law manner. Formally, this
statement does not contradict to the papers \cite{1}--\cite{5}.
Indeed, the Anderson theorem is valid under assumption of
a self-averaging character of the order parameter,
which in fact
reduces to its spatial uniformity. According to estimates of
\cite{3,4},
the self-averaging property
tends to violate when the
localization threshold is approached and the space-inhomogeneous
superconductivity is expected in the deep of the localization
phase; so the true  $T_c$  can be greater than its value
given by the Anderson theorem. In fact, controversy between
the papers  \cite{1}--\cite{5}  and \cite{9,10} has an
ideological character. The authors of \cite{1}--\cite{5} proceed
from the standpoint that localization counteracts to
 superconductivity, so the latter encounters a lot
of problems in the localized phase \cite{2,5}.
Contrary,
the growth of $T_c$ after the mobility edge \cite{9,10}
indicates
 that superconductivity not
only "survives" but even "prospers" in the localized phase. It
looks  suspicious
from the physical viewpoint and
contradicts the experimental situation, which is in complete
agreement with \cite{1}--\cite{5}.

The present paper has an aim to clarify a situation. In
fact, the essence of the problem is:
how and in what extent
 self-averaging
of the order parameter can be
violated?  The efficient approach to such problems was
developed in \cite{11}--\cite{14} and consists in the study
of individual defects and their  influence on the transition
temperature.
In particular,  for the plane
defects arranged perpendicularly to the $z$ axis
with the period $L$ along it, the change of
$T_c$ is determined by the formula\,\footnote{\,We accept for
simplicity that a plane defect changes only the density of
states. Generalizations of  (1), accounted for the
change of the interaction constant  $g$ \cite{12,13} and the
cut-off frequency  $\omega_0$ \cite{14} are also available.}
$$
\frac{\delta T_c}{T_{c0}} =
\frac{1}{\lambda_0^3 L}\, g^2 \!  \int dz \left[ \nu_0  \nu_1(z)
+  \nu_1(z)^2 \right]\,,
\eqno(1)
$$
if there are no surface states localized near defects. Here $g$ is
the Cooper interaction constant,  $\nu_1(z)$ is a deviation of the
local density of states $\nu(z)$ from its unperturbed value
$\nu_0$,  $\lambda_0=g\nu_0$ is the dimensionless coupling
constant,  $T_{c0}$ is the transition temperature in the absence
of defects, integration is carried out over a vicinity of the
single defect. For weak defects, only the linear in  $\nu_1(z)$
term is essential, which exactly corresponds to the Anderson
theorem and relates the change in  $T_c$ with the change of the
average density of states. Generally,  $ \nu_1(z)$ is comparable
with $\nu_0$ and already Eq.1 predicts a possibility of essential
violation of the Anderson theorem. It is related with the fact
that the initially uniform order parameter is influenced by
strong defects and can increase or decrease in their vicinity.
More essential violations of the Anderson theorem
are possible, if
the surface states localized near defect
appear
at the Fermi level (Fig.1).
\begin{figure}
\centerline{\includegraphics[width=2.5 in]{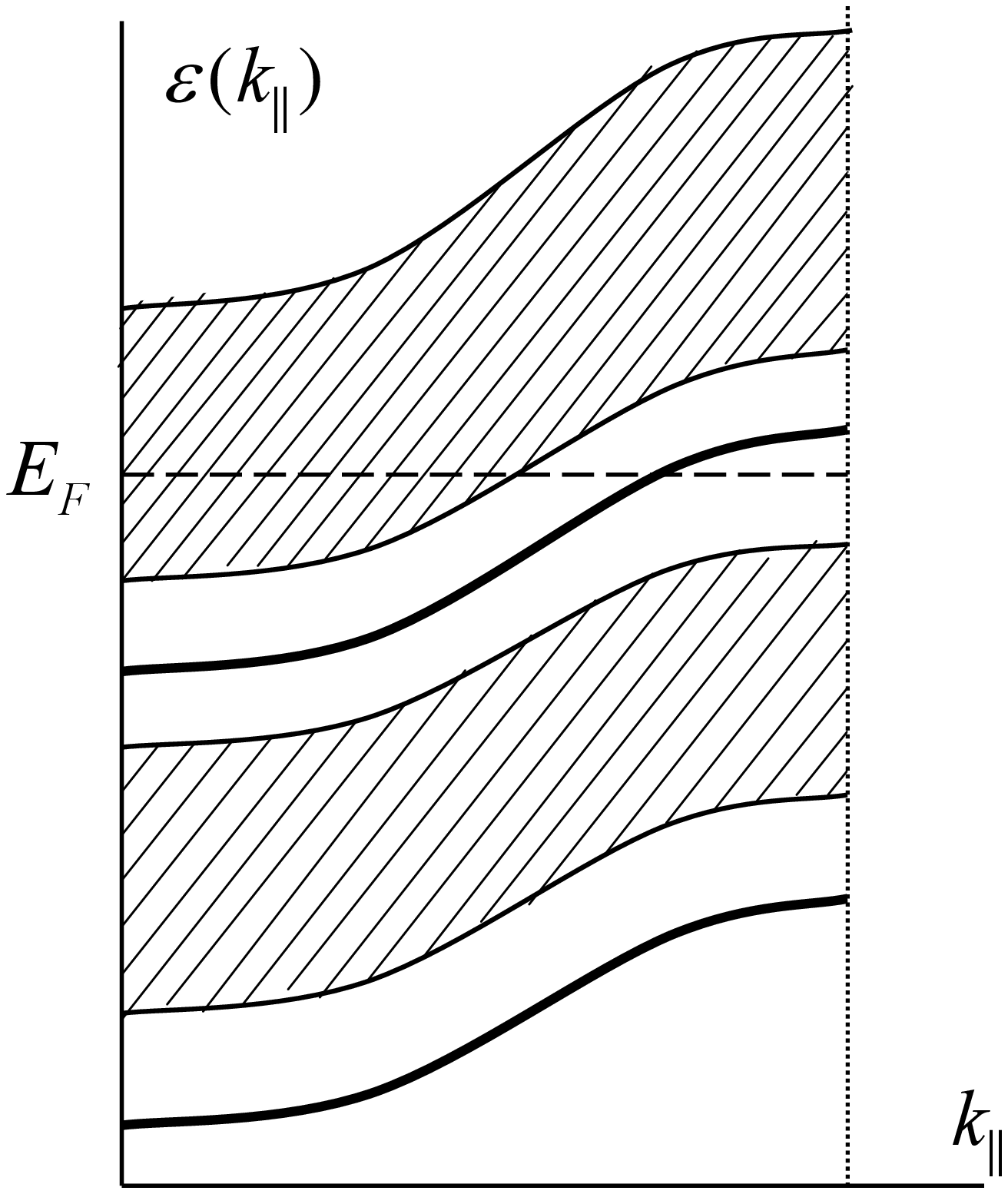}}
\caption{\footnotesize If plane defects are arranged perpendicular to
the  $z$ axis, the problem allows a separation of variables and
the longitudinal quasi-momentum  $k_\parallel$ can be introduced.
For fixed $k_\parallel$, the spectrum is a set of bands with
discrete levels splitted from them; if dependence on
$k_\parallel$ is taken into account, these levels turn into $2D$
bands, which can appear at the Fermi level.
} \label{fig1}
\end{figure}
In this case, the order parameter can
be strongly localized near the plane defects, so $T_c$ does not
depend on $L$ and is determined by the BCS formula $T_c=1.14
\omega_0 \exp(-1/\lambda_{2D})$ with the coupling constant
 $\lambda_{2D}$, corresponding to the separated
 two-dimensional band  (Fig.1). A crossover between two
 regimes is appeared to be very sharp and the intermediate
 situation is of little interest. Formally, the described results
 correspond to the periodical arrangement of defects, but their
 character shows that the assumption on periodicity is not
 essential; so they give a complete picture for the
 small "impurity" concentration in the 1D geometry.

Analogous effects are possible
in case of the point defects,
where the localized regime for the order parameter is related
with existence of the quasi-local states (Fig.2). A detailed
investigation of these effects allows to obtain the complete
picture of  possible violations of self-averaging.
The main conclusion is that such
violations are determined by  individual
defects and have no direct relation to the Anderson transition.
One can distinguish two typical situations.
\begin{figure}
\centerline{\includegraphics[width=2.5 in]{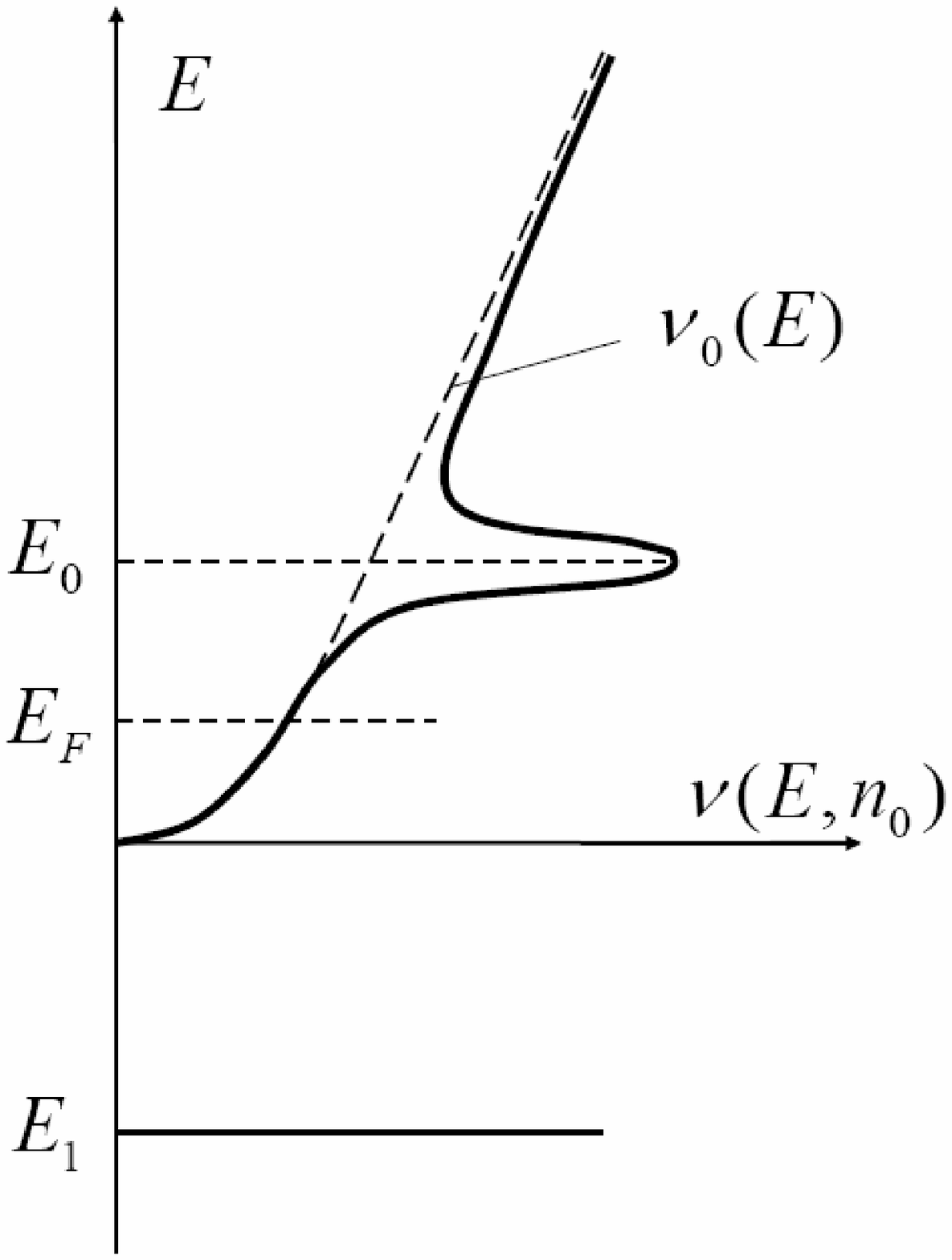}}
\caption{\footnotesize A strong point defect inserted  into
site $n_0$ of an ideal lattice
leads to appearance of
the local ($E_1$) and quasi-local  ($E_0$) levels. The latter
corresponds to the maximum of the local density of states
$\nu(E,n_0)$. } \label{fig2}
\end{figure}

If disorder is created by weak impurities (Fig.3,\,a), then
\begin{figure*}
\centerline{\includegraphics[width=5.0 in]{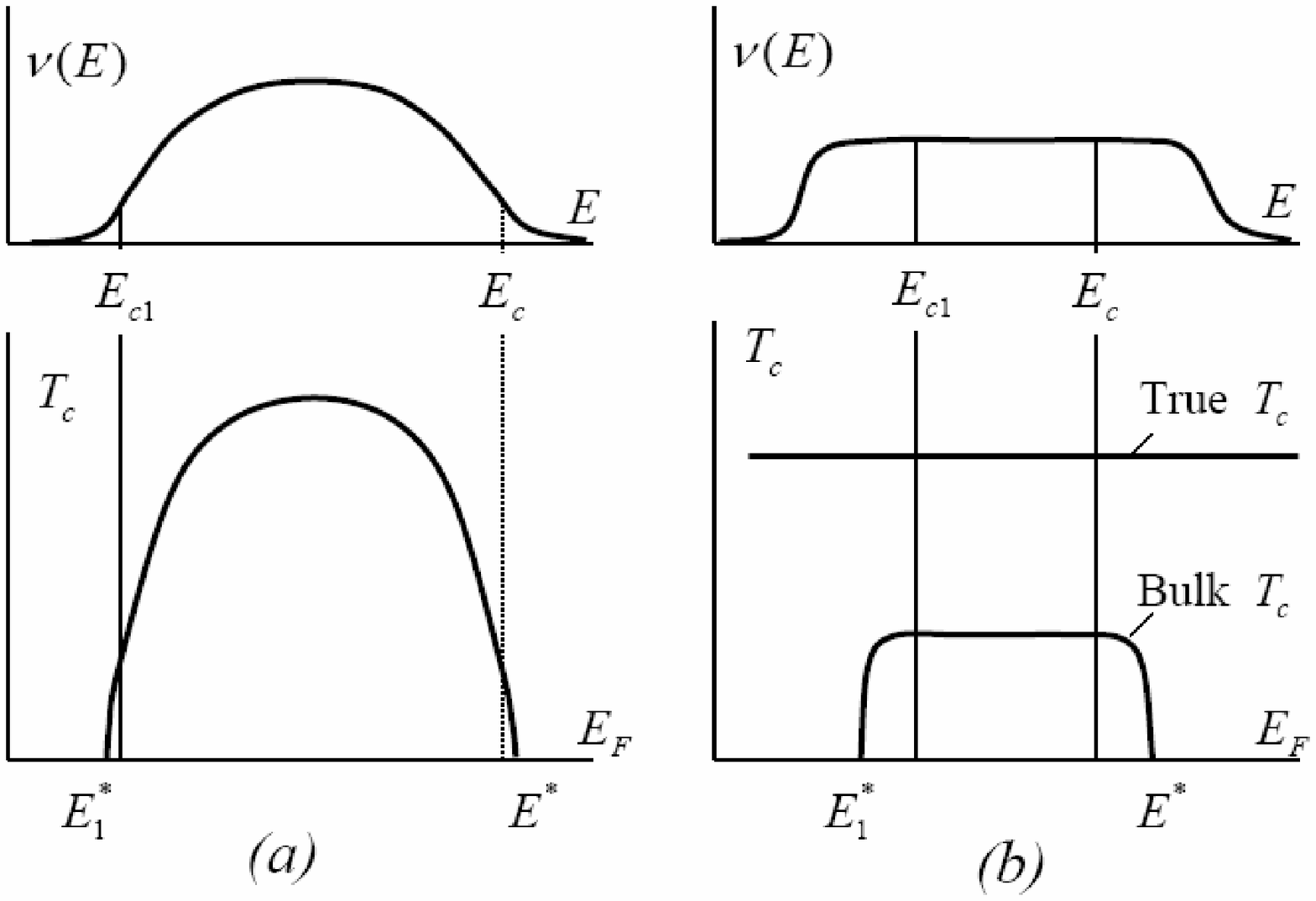}}
\caption{\footnotesize (a) If  disorder is created by weak
impurities, self-averaging of the order parameter is not
violated; $T_c$ strongly falls near the mobility edge due
to decrease of the density of states.  (b) In  case of strong
disorder, the true $T_c$ is determined by localization of the
order parameter at rare resonant impurities, while the
"bulk $T_c$" obeys to the  "rectangular" dependence. }
\label{fig3}
\end{figure*}
the assumption of self-averaging is always true and the
Bulaevskii--Sadovskii picture is literally applicable. The
mobility edge lies near the initial band edge and $T_c$ is
falling quickly at approaching  it from the metal side due to
decrease of the density of states; hence,
superconductivity
becomes practically unobservable before the mobility edge is
reached. Such situation is typical for the traditional
superconductors, which are good metals and effectively screen any
impurity which is introduced in them.  The experimental situation
is in complete agreement with these considerations \cite{5}.

In case of strong disorder  (Fig.3,b), the mobility edge can
be located in the region of the practically uniform density of
states\,\footnote{\,According to results by Zharekeshev \cite{51}
for the strongly disordered Anderson model, there is a wide
plateau for the density of states in the center of band.  }, so
the $T_c$ value given by the Anderson theorem does not fall in
approaching the localization threshold. In fact, the true  $T_c$
appears to be much larger and corresponds to localization of the
order parameter at the small number of the "resonant"
impurities, which
produce
the quasi-local states near the Fermi
level. In accordance with papers \cite{9,10}, $T_c$ depends on the
interaction constant $g$ in the power-law manner, but contrary to
them, it  has no essential dependence on the Fermi level
position.  It removes an illusion that localization
"helps"  superconductivity. In the vicinity of the true  $T_c$,
observation of superconductivity is practically impossible due to
a small fraction of the Meissner phase and negligible values of
the critical current.
Superconductivity becomes easily observable when it spreads to
the whole volume: it occurs at some effective temperature, which
we refer as the "bulk $T_c$"; it can be defined theoretically as a
transition temperature of the system with removed "resonant"
impurities. Such "bulk $T_c$"  corresponds
 qualitatively (but not quantitatively)
to the Anderson theorem and the Bulaevskii--Sadovskii
picture is confirmed at this level.  In such a case,
$T_c$ obeys
the "rectangular" dependence
(Fig.3,b), which exponentially weakly deviates from the
horizontal line near the mobility edge $E_c$, and exponentially
weakly deviates from the vertical line near the endpoint  $E^*$
of superconductivity. Such situation is typical for high $T_c$
superconductors, where coexistence of localization and
superconductivity is easily observable \cite{5}.

The estimate for the true  $T_c$
$$
T_c \sim g a^{-d} \sim \lambda_0 J
\eqno(2)
$$
($a$ is the lattice spacing, $J$ is a bandwidth, and $d$ is the
space dimension)
gives an impression
that the "room" superconductivity is a
widespread phenome- non.  In fact, the growth of  $T_c$ with the
increase of $\lambda_0$ is bounded by a quantity  $\omega_0/\pi$,
where $\omega_0$ is a cut-off frequency. For the phonon mechanism,
such upper bound  corresponds to the values already
attained in
high $T_c$ superconductors, and their further increase
requires the use of higher frequency Bose excitations. In
addition, the observation of true  $T_c$ is probably possible
only with the use of the scanning tunnel or squid microscopy
 \cite{50}.

It should be  stressed that Eq.2 is a result of  the mean
field theory. The corresponding solution for the order
parameter shows existence of the certain uniform
contribution with abrupt peaks
near the rare resonant impurities
(with concentration $T_c/J$). The order parameter can be
considered as positive (see Sec.2) and so its phase is
constant in the whole volume. In the fluctuational theory,
the modulus of the order parameter remains practically
unchanged, while the essential phase fluctuations arise.
If the uniform contribution is neglected, then the system
is divided into practically independent superconducting
"drops", whose phases are fluctuating freely and destroy
the macroscopical coherence of the superconducting state.
If the uniform contibution is taken into account,
the Josephson coupling between drops arises and
their phases become correlated. The accurate fluctuational
analysis of such a system is nontrivial, but the general
character of results is the same as for the
granual superconductors \cite{70}. If the ratio $T_c/J$ is not
too small, then the resonant impurities are close to each other
and their Josephson interaction is strong enough for
stabilization of the mean-field solution at  practically
the same  $T_c$  value (in this sense it can be qualified as
"true");  if a concentration of the resonant impurities appears
to be  small, then  $T_c$  is suppressed by fluctuations
to the value somewhat greater than the "bulk $T_c$" (Sec.\,7).

According to the results of  \cite{9,10}\,\footnote{
\,In the recent paper by Burmistrov et at \cite{71}
the results analogous to  \cite{9,10} are obtained
in the Finkelstein renormalization group approach
 \cite{72}. However, these papers are essentially
different both in the initial assumptions and in the discussed
physical mechanism, so one cannot say that one paper confirms
another.  The authors of \cite{9,10} tried to advance
beyond the assumption on self-averaging, while a fixed value of
the interaction constant is accepted; contrary,
  \cite{71} takes into account
a disorder dependence of the interaction constant, while
a self-averaging property is taken for granted. By the
latter reason, the present results cannot be reproduced in
\cite{71}, whereas the considered there effect  is more weak.
}
$$
T_c \sim g^{1/\gamma}\,, \qquad f\sim (T_c/J)^\gamma\,,
\eqno(3)
$$
where  $f$ is a portion of volume occupied by superconductivity,
and parameter
$\gamma=0.57$ is related with a fractal dimensionality of wave
functions. We do not deny the existence of the order parameter
configurations, leading to  results of type  (3) (Sec.3), but
Eq.2 corresponds to the higher value of $T_c$;
the corresponding configuration of the order parameter
is determined by the rare peaks near the resonant
impurities, occuring at the atomic scale and occupying
a portion of volume  $f\sim T_c/J$. If
superconductivity is considered as a variational problem,
then it is possible to  say that our trial function is
more successful than one in \cite{9,10}.
Formally, our results correspond to Eq.3 with
$\gamma=1$ and do not contain any information on multifractality;
hence, there are no grounds to say on "fractal superconductivity"
\cite{10} near the localization threshold.

\begin{center}
{\bf 2. Anderson theorem and inequalities for  $T_c$ }
\end{center}

A basis for description of the spatially inhomogeneous
superconductivity is given by the Gor'kov equation for
the order parameter   $\Delta(r)$
$$
\Delta(r) =  \int  K ( r,  r')
\Delta( r') d^d r'
\eqno(4)
$$
with the kernel  $K(r,r')$ in  representation of exact
one-particle eigenstates  $\varphi_s(r)$
$$
K(r,r')=  g T \sum \limits_{\omega} \sum\limits_{s,s'}
\frac{\varphi^*_s(r) \varphi_s(r') \varphi^*_{s'}(r)
\varphi_{s'}(r')} {(\epsilon_s-i\omega)(\epsilon_{s'}+i\omega)}
\,,
\eqno(5)
$$
where  $\epsilon_s$ are eigenenergies (counted from the Fermi
level), and summation occurs over the Matsubara frequencies
$\omega_n=\pi T(2n+1)$ with integer  $n$.
Following  de
Gennes \cite{15}, we use the frequency cut-off
$|\omega|<\omega_0$, which corresponds to the electron interaction
$$
V(r,r'; \omega)= - g\, \theta(\omega_0-|\omega|)
\,\delta (r-r')\,,
\eqno(6)
$$
which is strictly local and can be specified independently of
one-particle eigenstates
(in contrast to the momentum cut-off in the
original BCS formulation, where interaction is defined by the
matrix elements over plane waves). In the absence of  magnetic
effects, eigenstates $\varphi_s(r)$ can be taken real and their
orthogonality leads to the sum rule  \cite{15}
$$
\int  K (r,r') d^d r' = g \nu_F(r)
\ln\frac{1.14\omega_0}{T} \,,
\eqno(7)
$$
where  $\nu_F(r)\equiv \nu(0, r) $ is the local density of
states
$$
\nu(\epsilon, r)= \sum\limits_{s} |\varphi_s(r)|^2 \,
\delta(\epsilon -\epsilon_s)
\eqno(8)
$$
at the Fermi level.
It is accepted in derivation of  (7) that
$\nu(\epsilon, r)$ is a slow function of  $\epsilon$ on
the scale of $T_c$; generally  $\nu_F(r)$
should be understood as a local density of states smoothed
at the scale of  $T_c$.

The Anderson theorem follows from Eq.4 under assumption
of a self-averaging
 order parameter, when
$\Delta(r)$ and $K(r,r')$  can be independently averaged
over disorder. Since  $\langle\Delta(r) \rangle$ does not
depend on  $r$ due to the spatial uniformity in average, the use
of the sum rule  (7) gives
$$
\langle\Delta\rangle= g \langle\nu_F\rangle
\ln\frac{1.14\omega_0}{T} \langle\Delta\rangle \,,
\eqno(9)
$$
and $T_c$ is given by the BCS formula, which contains the
average density of states $\langle\nu_F\rangle$. The latter
does not change at the Anderson transition point, suggesting
the analogous behavior for  $T_c$. More detailed information can
be obtained,
if Eq.4 is averaged over variable  $r$
$$
\langle\Delta\rangle= g
\ln\frac{1.14\omega_0}{T} \langle\nu_F(r)\Delta(r)\rangle \,.
\eqno(10)
$$
The function  $\Delta(r)$ can be considered as
positive\,\footnote{\,For real $\varphi_s(r)$, the kernel
$K(r,r')$ is positive, since it  can be written as  $g T
\sum_{\omega} \left|G_\omega(r,r')\right|^2$ (see Eq.37). The
Cooper instability corresponds to the minimal characteristic
number (or maximal eigenvalue)
and  the nodeless eigenfunction (the Entch
theorem) \cite{18}.}, and one has
$$
\nu_{min}\,\langle\Delta\rangle
\le \langle\nu_F(r)\Delta(r)\rangle \le
\nu_{max}\,\langle\Delta\rangle  \,,
\eqno(11)
$$
where $\nu_{min}$ and  $\nu_{max}$ are the minimal and maximal
values of $\nu_F(r)$.  It gives inequalities for $T_c$
$$
1.14\omega_0 e^{ -1/g\nu_{min} } \le T_c \le
1.14\omega_0 e^{ -1/g\nu_{max} }                              \,,
\eqno(12)
$$
which can be also obtained from the known theorems of the
matrix theory \cite{13}[Sec.2].  According to Eq.12,
the power law dependence of  $T_c$ on the coupling constant
$g$ \cite{9,10} is impossible, if $\nu_F(r)$ has an
upper bound  $\nu_{max}$.

Near the Anderson transition, there are systematic reasons
for  growth of the $\nu_F(r)$ fluctuations \cite{3,4}. As
noted in  \cite{5}, the correlator  $\langle
\nu(E+\omega,r)\,\nu(E,r')\rangle$ at $r=r'$ coincides
with the Berezinskii--Gor'kov  spectral density
\cite{300}, which is determined by the diffusion pole
with the observable diffusion coefficient $D(\omega,q)$
\cite{301}:
$$
\langle \nu(E+\omega,r)\,\nu(E,r)\rangle
\sim {\rm Re} \int \frac {d^dq}{-i\omega + D(\omega,q) q^2}\,.
\eqno(13)
$$
In the metallic phase, the static diffusion constant  $D(0,q)$
is real, so  $\langle \nu_F(r)^2\rangle$ diverges at the
transition point as  $ D^{-1}$.  In the dielectric phase,
the analogous estimate can be obtained from  the
self-consistent theory of localization  \cite{302} by
iteration of Eq.112 in \cite{301}
$$
D(\omega,q)= (-i\omega) d(q) +\omega^2 d_1(q)\,,
$$
$$
d(q)\sim \xi^2\,,
\quad  d_1(q)\sim \xi^4 \,|\tau|^{-1}
\eqno(14)
$$
($\tau$ is a distance to the critical point), so
$\langle \nu_F(r)^2\rangle \sim |\tau|^{-1}$ and
fluctuations grow symmetrically on two sides of the
transition\,\footnote{\,According to the self-consistent
theory,  $D\sim \tau$  in the metallic phase \cite{302}.}.
Estimations of the correlator (13) at the critical point
based on  multifractality  of wave functions  \cite{10}
suggest the dependence $\omega^{-\gamma}$ for  $\omega\to
0$; if divergency is  cut off  at the scale  $T_c$, then
 $\nu_{max}\sim \nu_0 (J/T_c)^{\gamma/2}$ and the maximum
value $T_c\sim g^{2/\gamma}$ allowed by  Eq.12 is in a
qualitative agreement with \cite{9,10}.
Consequently, if the upper bound for  $T_c$ is realized
in Eq.12, then it reaches the maximum value at the transition
point
depending on  $g$ in the power law manner.

However, the distribution of  quantities  $\varphi_s(r)$
has the power law tails \cite{10} and Eq.13  determines
neither the typical, nor the maximal value of $\nu_F(r)$.
In fact, the given estimate for $T_c$ is not reached for weak
disorder and is exceeded for strong disorder. Formally, the
approach of  \cite{10} is
questionable due to replacement of matrix
elements $M_{ijkl}=\int d^dr
\varphi_i(r) \varphi_j(r) \varphi_k(r) \varphi_l(r)$ by their
mean values with averaging independently of the order
parameter.

More efficient approach is based on the study of effects
from individual impurities, since it allows to work with
specific realizations of the random potential and contains no
problems of averaging. Introducing one impurity after another,
one can easily be convinced (Sec.4), that unbounded values of
 $\nu_F(r)$ can arise only from existence of quasi-local states
(Fig.2). The problem of quasi-local states has a general
character. Indeed, one can imagine such fluctuation of the random
potential, that a  finite region of space is isolated from its
environment by the high barrier; the corresponded discrete levels
can have a very weak broadening and, appearing close to the Fermi
level, can lead to unbounded values of $\nu_F(r)$. Such problems
are discussed in the next section.

\begin{center}
{\bf 3. Resonances at quasi-discrete levels } \end{center}

Suppose that a system has a discrete
spectrum and only one state is close to the Fermi
level; then we can retain only one term in the sum over $s,\,s'$
in (5):
$$
K(r,r')= g T\, \sum \limits_{\omega}
\,\frac{\varphi^2_0(r) \,\varphi^2_0(r')}
{\epsilon^2_0+\omega^2}  \equiv
$$
$$
\equiv g A(T)
\,\varphi^2_0(r) \,\varphi^2_0(r') \,,
\eqno(15)
$$
Then Eq.4 gives
$$
\Delta(r) = X \varphi^2_0(r),
$$
$$
 X=g A(T) \int
\varphi^2_0(r') \Delta( r') d^d r'\,. \eqno(16)
$$
and  self-consistency of these equations
determines  $T_c$:
$$
1 =g A(T) I_4\,, \qquad I_4= \int \varphi^4_0(r) d^d r \,.
\eqno(17)
$$
Calculation of  $A(T)$ is possible without the cut-off
frequency taken into account, since the sum converges
at large $\omega$:
$$
A(T)= T \sum \limits_{\omega}
\frac{1}{\epsilon^2_0+\omega^2} =\frac{1}{2\epsilon_0}
\tanh\frac{\epsilon_0}{2T}   \,.
 \eqno(18)
 $$
For the exact resonance  ($\epsilon_0=0$) we have
$A(T)=1/4T$, so
$$
 T_c=g I_4/4   \,,
 \eqno(19)
$$
and  $T_c$ has a power law dependence on  the  interaction
constant $g$. In the general case  (see Fig.4,a)
\begin{figure*}
\centerline{\includegraphics[width=5.1 in]{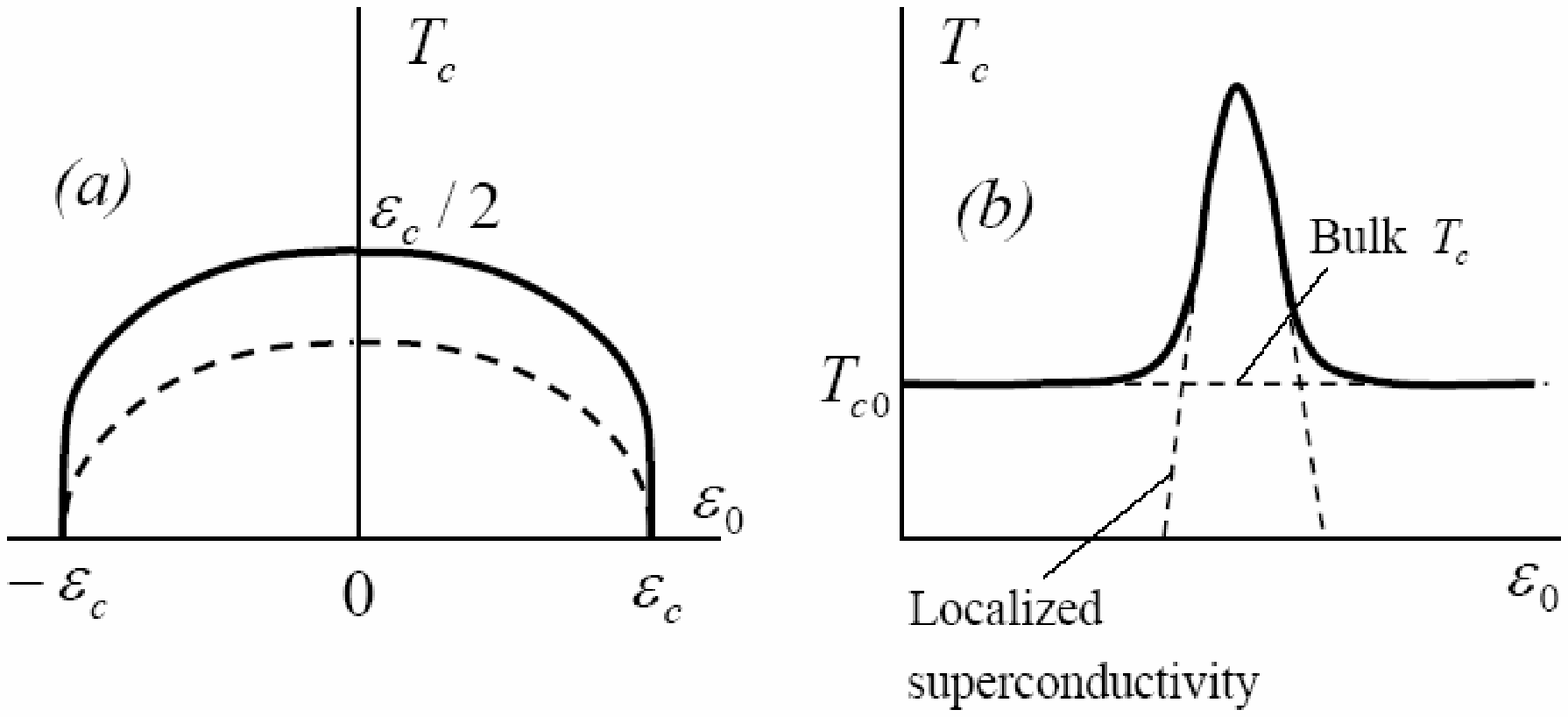}}
\caption{ \footnotesize (a) $T_c$ of the one-level system as  a
function of the level position $\epsilon_0$ in the absence of
attenuation (solid line); attenuation $\gamma$ produces
a shift of the curve by the quantity  $\sim \gamma$
(dashed line).  (b) The same, for a situation when the
quasi-discrete level lies in the background of the continuous
spectrum.} \label{fig4}
\end{figure*}
$$
T_c= \frac{\epsilon_0}{\ln(\epsilon_c+\epsilon_0)
-\ln(\epsilon_c-\epsilon_0)} \,,\qquad \epsilon_c=gI_4/2 \,,
 \eqno(20)
 $$
and a solution exists under condition
$$
|\epsilon_0|<\epsilon_c  \,.
 \eqno(21)
$$
At first glance, the considered regime is destroyed due to
fluctuations\,\footnote{\,If eigenstate $\varphi(r)$
is localized, then according to (20) a superconducting
transition
takes place in a finite system;
of course, such conclusion is an artifact of the mean field
theory and in fact the transition is destroyed by fluctuations. }
or
coupling
with the continuous
spectrum; in fact, it is not so (see below) and the main problem
consists in the possibility to match the discrete level with the
Fermi energy.

Indeed, let the system has a finite size  $L$, while its
eigenstates are extended. Then the Fermi energy is
located between two discrete levels\,\footnote{\,For a
discussion of the parity effect see Footnote 18.}, and
$\epsilon_0$ is determined by the average level
spacing $J (L/a)^{-d}$; estimating $I_4\sim L^{-d}$ from the
normalization condition, we see that
$$
\epsilon_0 \sim J (L/a)^{-d}\,, \qquad \epsilon_c \sim g L^{-d}
 \eqno(22)
$$
and  condition (21) cannot be fulfilled in the weak coupling
regime, which is the only allowable in the BCS scheme.

Let us couple the given system with a reservoir,
and try to match the chemical potential of the latter with
the discrete level of the system. However, nothing good
will occur from it: the local Fermi level of
the system is still arranged between two discrete levels
and it tends to
equalize with the Fermi energy in the reservoir. The real flow of
electrons is impossible due to elecroneutrality, and the problem
will be solved by a minimal deformation: a double layer will arise
between the reservoir and the system, and it will equate the
Fermi levels.

By the same reason, the situation cannot be improved due to
localization of states. At first glance, in this case $\epsilon_0
\sim J (L/a)^{-d}$, $\epsilon_c \sim g \xi^{-d} $ ($\xi$ is the
localization radius of $\varphi_0(r)$), so  condition (21)
reduces to $g \agt  J a^d (L/ \xi)^{-d} $ and can be fulfilled at
sufficiently large $L$. In fact, blocks of size  $\xi$ are
quasi-independent and each of them has its own local Fermi level;
these levels equalize due to double layers between blocks,
and the given estimates are valid only for $L\sim \xi$. In fact,
the above arguments clarify the mechanism for the Coulomb gap
\cite{19}.

It looks that the only possibility to avoid the given
arguments is to take the size  $L$ of the atomic order.
Indeed, at such a scale: (a) the notion of the Fermi level
becomes senseless; (b) electroneutrality can be
violated; (c) a size of the double layer is comparable with
$L$. It means that
the strong violations of the
Anderson theorem can be  exhaustively
analyzed by consideration of the one-impurity problem (Sec.4).

Already at this stage it is possible to establish the relation
with results of \cite{9,10}. In
the considered
there strictly one-electron picture, the discrete system of
levels fluctuates freely relative to the Fermi energy,
so resonances are possible at any length scale  $L$. Then all
principal statements of \cite{9,10} are reproduced: $T_c$ has a
power law behavior as a function of  $g$ and does not depend on
the cut-off frequency $\omega_0$, while the order parameter
$\Delta(r)$ follows the form of the wave function (see (16)) and
will have multifractal properties simultaneously with
multifractality of the latter.\,\footnote{\,We have no doubt that
papers \cite{9,10} implicitly dealt with the same effect, but the
improper averaging procedure led to
a domination  of large
length scales.  } However, this picture is completely
destroyed, if electroneutrality is taken into account,
since resonances at large scales become impossible.
In fact, large scale fluctuations are insignificant even in
a strictly one particle picture: a value of $T_c$ for an exact
resonance, $T_c\sim g L^{-d}$ (see(17),(19)), is greater for
small scales.

Generally, the considered regime is not destroyed in the presence
of the continuous spectrum. In this case, the level $\epsilon_0$
acquires the finite decay  $\gamma$, which can be
taken into account by replacement
$$
\epsilon_0\pm i\omega\, \,\longrightarrow  \, \,
\epsilon_0\pm i\omega \pm i\gamma\, {\rm sign}\, \omega
\eqno(23)
$$
so
$$
A(T)= T \sum \limits_{\omega}
\frac{1}{\epsilon^2_0+(|\omega|+\gamma)^2}
\approx
$$
$$
\approx \frac{1}{\pi\epsilon_0}
\arctan\frac{\epsilon_0}{\gamma+bT} \,,
 \eqno(24)
 $$
where we have estimated the sum by the integral, introducing
the cut-off  $|\omega|>bT$ (for the choice $b=4/\pi$ such
estimate practically coincides with the exact result (18)
for $\gamma=0$). The finiteness of  $\gamma$ leads
qualitatively to the shift of the curve in Fig.4,a by a
quantity  $\sim\gamma$, so a solution survives
for  $\gamma\alt \epsilon_c$.

For finite  $\omega_0$ one obtains instead  (24)
$$
A(T) \approx \frac{1}{\pi\epsilon_0}
\arctan\frac{\epsilon_0 (\omega_0 -bT)}
{\epsilon_0^2+(\omega_0+\gamma)(\gamma+bT)} \,,
 \eqno(24')
 $$
and can be easily convinced that finiteness of $\omega_0$
is irrelevant under condition $\omega_0\gg \epsilon_c$.  In
the opposite case  $\omega_0\ll \epsilon_c$ the allowed
values of  $\epsilon_0$ and $\gamma$ have an order
$( \epsilon_c \omega_0)^{1/2} $, while the maximal critical
temperature $T_c$ is of the order $\omega_0$; in fact,
restriction  $T_c < \omega_0/\pi$ is evident, since
for  $T>\omega_0/\pi$  the sum over  $\omega$ contains no
terms.

To investigate the effect of the continuous spectrum
on the order parameter, one can use
the following approximation for the kernel   $K(r,r')$
$$
K(r,r')= K_0(r-r')+ g A(T)
\varphi^2_0(r) \varphi^2_0(r') \,,
\eqno(25)
$$
which ignores the backward influence of the discrete level
on the continuous spectrum. According to  \cite{11,12}, such
approximation provides qualitatively correct description
and can be justified in certain limiting cases.\,\footnote{\,In
Sec.4 we consider the one-impurity problem with the backward
influence on the continuous spectrum.}

Having in mind a consideration of periodical configurations,
we solve Eq.4 with the kernel  (25) for a finite system of
size  $L$ with the periodical boundary conditions. We
accept $L\ll \xi_0 \tau^{-1/2}$, where
$\tau=(T-T_{c0})/T_{c0}$, $\xi_0$ is the coherence length,
and $T_{c0}$  is a transition temperature, corresponding to
the continuous spectrum.\,\footnote{\,Appearance of
the characteristic scale
$\xi(T)=\xi_0 \tau^{-1/2}$ was discussed previously \cite{11}
for the case of plane defects. If $L\gg \xi(T)$, then
individual defects becomes practically independent and the order
parameter is localized near them on the scale  $\xi(T)$. In the
opposite case $L\ll\xi(T)$, the order parameter is practically
constant in the space between defects. Below (Sec.5) we
consider
configurations with small concentration  ($\sim T_c/E_F$)
of the resonant impurities, so the distance between them
 $a (E_F/T_c)^{1/3}$  is less than $\xi_0\sim a (E_F/T_c)$.
}
After the Fourier transform one has
$$
\Delta_q = gA(T) X
\frac{\langle\varphi^2_0 \rangle_q }{1-K_0(q)} \,,
$$
$$
X=L^{-d} \sum\limits_q \langle\varphi^2_0 \rangle_{-q}
\Delta_q
\eqno(26)
$$
and  self-consistency of two expressions  leads to
$$
1=g A(T) \left[ L^{-d} \sum\limits_q
  \langle\varphi^2_0\rangle_{q} \langle\varphi^2_0\rangle_{-q}
+ \right.
$$
$$
\left.  +L^{-d} \sum\limits_q \frac{K_0(q)}{1-K_0(q)}
  \langle\varphi^2_0\rangle_{q} \langle\varphi^2_0\rangle_{-q}
	   \right]   \,.
\eqno(27)
$$
Using expansion in   $q^2$
$$
1-K_0(q)= \lambda_0\tau +{\scriptstyle
\frac{1}{2}}\lambda_0\xi_0^2 q^2 +\ldots \,,
\eqno(28)
$$
it is easy to see that the main contribution to the
second sum in (27) occurs from  small  $q$, and the single term
with $q=0$ is sufficient for  $L\ll \xi_0 \tau^{-1/2}$.  Then
Eq.27 accepts a form
$$
1=g A(T) \left[ I_4+\frac{1}{\lambda_0 L^d \tau} \right]\,,
\eqno(29)
$$
and the analogous approximations in  (26) give
$$
\Delta(r)=const \left[ \varphi_0^2(r)+\frac{1}{\lambda_0 L^d
\tau} \right]   \,.
\eqno(30)
$$
If  $T_{c0} \ll \epsilon_c$, then dependence of $T_c$ on
$\epsilon_0$ has a form shown in Fig.4,b.  In the zero
approximation there are two independent systems,
the quasi-local one
with the transition temperature (20) (if attenuation
 $\gamma$ is small) and
 the continuous one
 characterizing by  $T_{c0}$,
 while  $T_c$ of the composed system is given by the
 maximal of two values.
 Interrelation  of two systems
 reduces to smoothing  of dependence $T_c(\epsilon_0)$
at the scale  $T_{c0} (a/L)^{d/2}$, if  $\varphi(r)$ is localized
at the atomic scale  $a$.

It is clear from (30) that the order parameter  $\Delta(r)$
is practically constant for small $\tau$ and localized at the
scale $a$ for large  $\tau$. Crossover from one
regime to another is very abrupt, and one can say on
the "Anderson transition" for superconducting electrons.
We see that the localized regime survives in the presence of the
continuous spectrum, if the corresponding  $T_c$
exceeds  $T_{c0}$. In fact, existence of the continuous spectrum
has a stabilizing effect on the localized superconductivity, since
the order parameter takes non-zero values in the whole volume.

\begin{center}
{\bf 4. One-impurity problem }
\end{center}

If $G^0_{nn'}$ is the Green function of an ideal lattice,
$V_{n}=V\delta_{nn_0}$ is an impurity potential, then
the Green function $G_{nn'}$ of the perturbed system is
determined by the Dyson equation \cite{21}:
$$
G_{n n'}=G^0_{n n'}+G^0_{n n_0}V G_{n_0 n'} \,.
\eqno(31)
$$
Setting $n=n_0$, one has the closed equation for $G_{n_0 n'}$,
whose solution is substituted into (31)
$$
G_{n n'}=G^0_{n n'}+G^0_{n n_0} {\cal T} G^0_{n_0 n'} \,,
\qquad
$$
$$
{\cal T}=\frac{V}{1-V G^0_{n_0 n_0}}\,,
\eqno(32)
$$
where the scattering  ${\cal T}$-matrix  reduces
to a constant in the given case. For an ideal
lattice $G^0_{n n}$ does not depend on $n$,
$$
G^0_{nn}= \int \frac{d^dk}{(2\pi)^d} \frac{1}
{E-\epsilon_k+i 0} =
$$
$$ =
\int  \frac{\nu_0(\epsilon)\,d\epsilon }
{E-\epsilon+i 0}
\equiv I(E) -i \pi \nu_0(E) \,
\eqno(33)
$$
and so   ${\cal T}$-matrix has no $n_0$ dependence.
Condition $1-V I(E)=0$  corresponds to existence
of the local (if
$\nu_0(E)=0$) or quasi-local (if $\nu_0(E)\ne 0$) level
\cite{21,22}  (Fig.5). The local density at  site  $n_0$
\begin{figure*}
\centerline{\includegraphics[width=4.8 in]{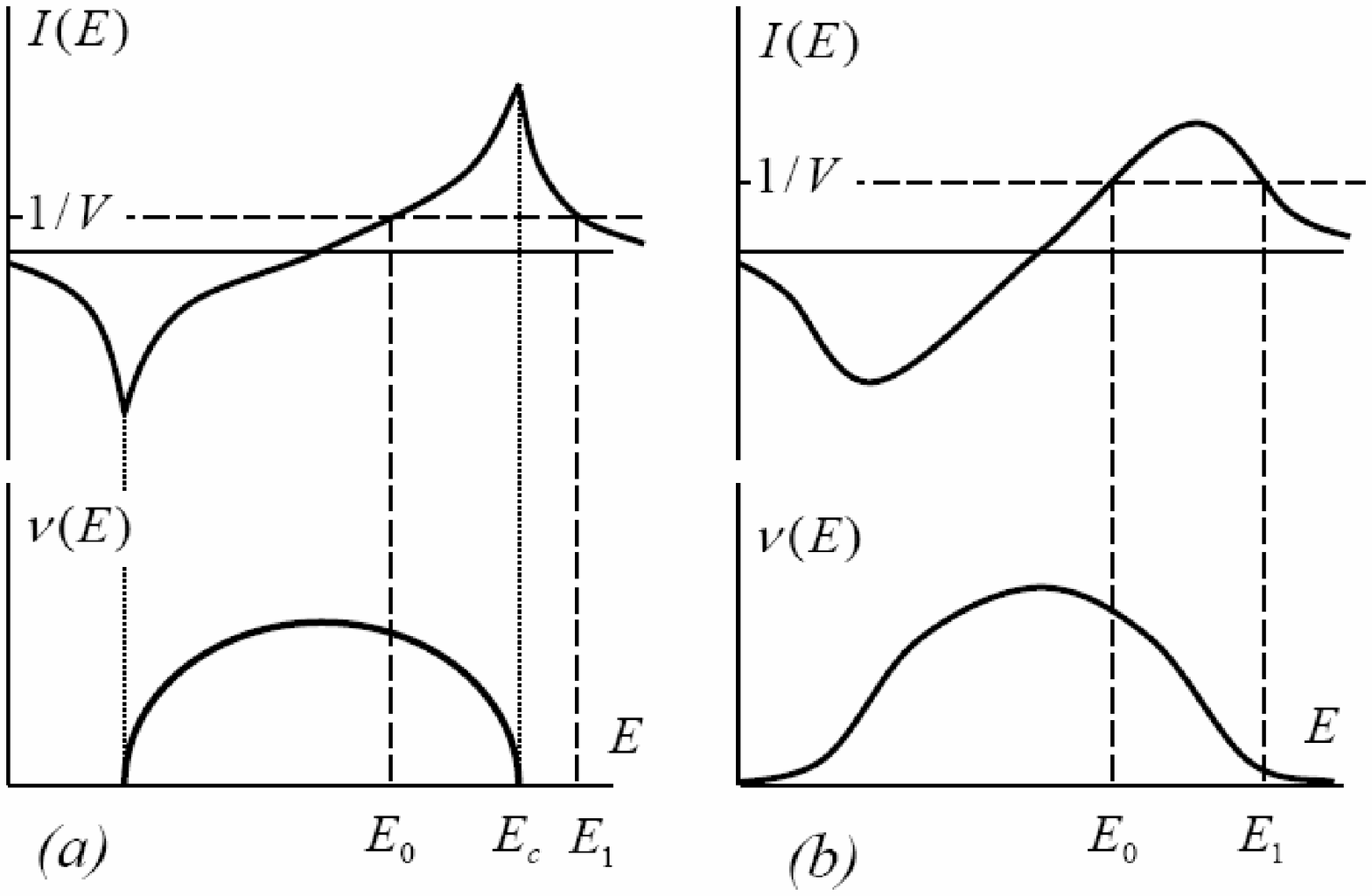}}
\caption{\footnotesize (a) If the impurity $V\delta_{n n_0}$ is
inserted in an ideal lattice, equation $1=V I(E)$ has
two roots for large $V$,  $E_1$ and  $E_0$, which
correspond to the local and quasi-local levels.  (b) If the same
impurity is inserted in the disordered lattice, both
solutions correspond to the quasi-local levels.} \label{fig5}
\end{figure*}
$$
\nu(E,{n_0})=  \frac{ \nu_0(E)}
{[1-VI(E)]^2 + [\pi V \nu_0(E)]^2}
\eqno(34)
$$
has an abrupt maximum near the resonance $1-V I(\epsilon_0)=0$
(Fig.6) with a value in it
\begin{figure}
\centerline{\includegraphics[width=3.1 in]{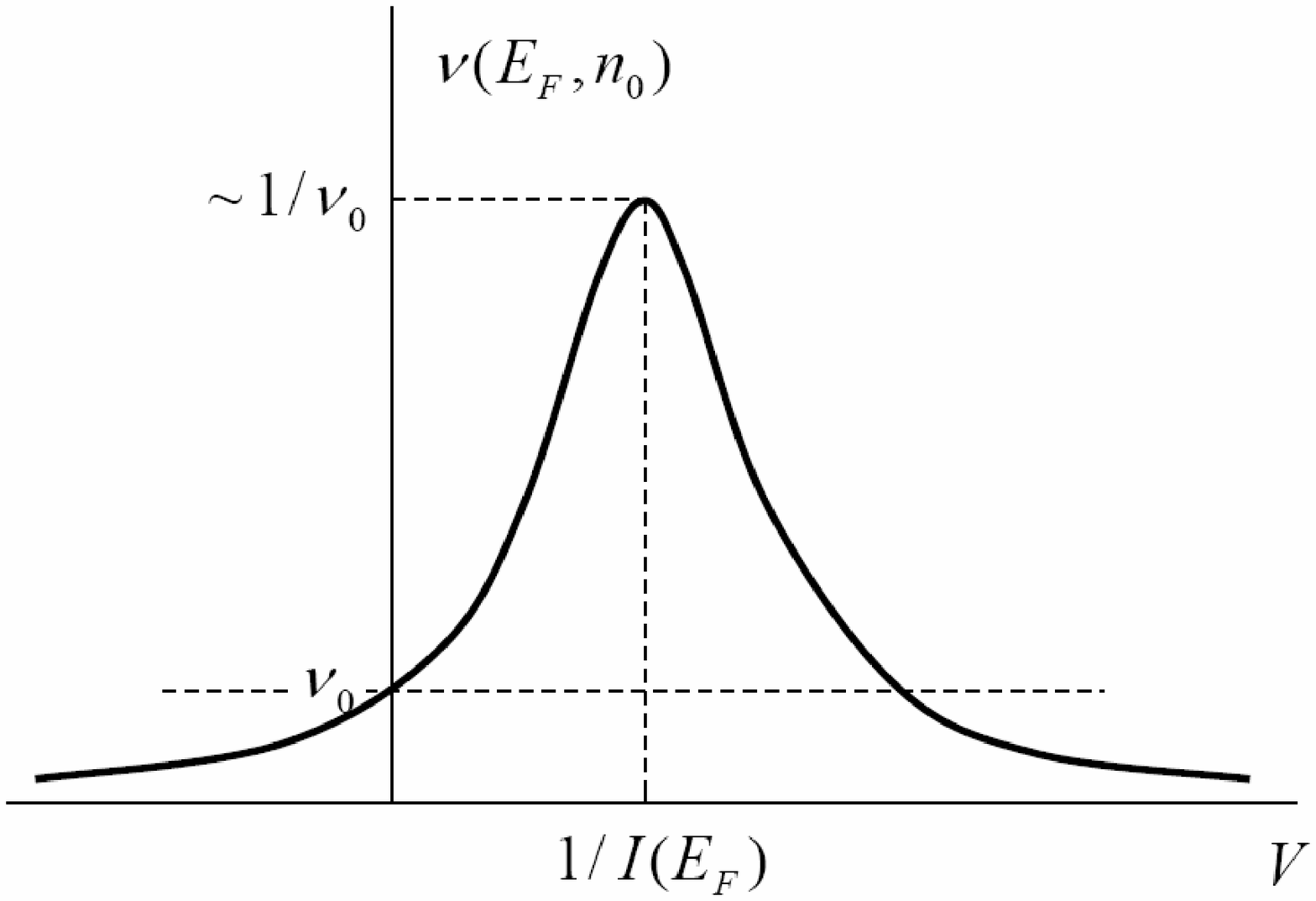}}
\caption{\footnotesize The local density of states at the
point  $n_0$ as a function of the impurity potential $V$. }
\label{fig6}
\end{figure}
$$
\left[\nu(E,{n_0})\right]_{res}= \frac{ I(\epsilon_0)^2}
{\pi^2  \nu_0(\epsilon_0)}
\eqno(35)
$$
which grows unboundedly  near the initial band edge (where
$\nu_0(\epsilon_0) \to 0$).  In the vicinity of the band edge, a
calculation of $G^0_{nn'}$ is possible in the continual
approximation and gives at $d=3$  (for $|r-r'| \agt a$ ):
$$
G^0_{nn'}\, \longrightarrow \, G^0(r-r') = -\pi \nu_0 \,
\frac{e^{ik_F |r-r'|}}{k_F|r-r'|} \,,
 \eqno(36)
 $$
Deviation of  $\nu(E,{n_0})$ from $\nu_0(E)$
is maximal for $n=n_0$ and tends to zero for
$|n-n'|\to \infty$.

The Matsubara representation for the Green functions is
obtained from  (32) by replacement
$E\to \epsilon_F +i\omega$, where  $\epsilon_F=0$ for the
corresponding choice of the energy origin. Setting
$n_0=0$, one has for the kernel in (4)
$$
K(r,r') = g T \sum_{\omega} \left|G_\omega(r,r')\right|^2=
$$
$$
=g T \sum_{\omega} \left|G^0_\omega(r-r')\right|^2 +
\qquad\qquad\qquad
$$
$$
\quad+gT \sum\limits_\omega\, G^0_\omega(r) \,
       {\cal T}_\omega \,G^0_\omega(r') \,G^0_{-\omega}(r-r')+
$$
$$
\qquad+ gT \sum\limits_\omega \,G^0_{-\omega}(r) \,
 \,{\cal T}_{-\omega}\, G^0_{-\omega}(r')\, G^0_{\omega}(r-r')+
$$
$$
+ gT \sum\limits_\omega\, \left|G^0_{\omega}(r)\right|^2\,
 \left|{\cal T}_{\omega}\right|^2 \,\left|G^0_{\omega}(r')\right|^2
\equiv     \quad
$$
$$
\equiv K_0(r-r')+ K_1 (r,r')
 \,. \eqno(37)
$$
Solution of the Gor'kov equation with the kernel  (37)  is sought
in the form
$$
\Delta(r) = \Delta_0 + \Delta_1 (r)    \,,
 \eqno(38)
 $$
where $\Delta_1 (r)$ is localized near $r=0$.
Substituting in  (4) and using the sum rule (7),
one has
$$
\Delta(r) =  \int  K_0 ( r-r') \Delta( r') d^3 r' +
$$
$$ +
g \nu_1(r) \ln\frac{1.14\omega_0}{T} \Delta_0 + F(r) \,
\eqno(39)
$$
where  $\nu_1(r)$ is  deviation of the local density of
states from  $\nu_0$ and
$$
F(r) =  \int  K_1 ( r,r') \Delta_1( r') d^3 r'  \,.
\eqno(40)
$$
Consideration of the isolated impurities is not actual
(see Footnote 6), so we accept their periodical
arrangement
and solve the Gor'kov equation for a finite system of size
$L$ with periodical boundary conditions for
$L\ll \xi_0 \tau^{-1/2}$.
Resolving (39) for  $\Delta(r)$ by the Fourier transform and
and simplifying the result analogously to  (27),
it is possible to
separate  the uniform term corresponding to $\Delta_0$,
while the rest is attributed to  $\Delta_1(r)$
($\langle\ldots\rangle_0$ is the zero Fourier component):
$$
\Delta_0 = \frac{K_0(0)}{\lambda_0 L^3 \tau}
 \left[\, \Delta_0 \ln\frac{1.14\omega_0}{T} \,
g\langle\nu_1\rangle_0 + \langle F\rangle_0
 \,\right]   \,,
$$
$$
\Delta_1(r) =  \Delta_0 \,\ln\frac{1.14\omega_0}{T} \,g\nu_1(r)
 +\, F(r) \,.
\eqno(41)
$$
Using the explicit expression for $F(r)$ and setting in the
integrals\,\footnote{\,This approximation is not quite
rigorous, but in fact it is used only for estimates:
the corresponding terms characterized by
parameters  $\lambda_{02}$ and $\lambda_{12}$ have no
significance both  far from the resonance, and in its
vicinity (see Appendix).}
$$
\Delta_1(r') G^0_{\omega}(r-r') \approx
\Delta_1(r') G^0_{\omega}(r)\,,
 \eqno(42)
 $$
one can transform (41) to the form
$$
\Delta_0 = \frac{K_0(0)}{\lambda_0 L^3 \tau}
\,\ln\frac{1.14\omega_0}{T} \,
 \left[  \vphantom{A^2_6}
  \, g\langle\nu_1\rangle_0 \,\Delta_0  \,
 +\, g\langle\nu_1 \Delta_1\rangle_0  \,\right]   \,,
\eqno(43a)
$$
$$
\Delta_1(r) =  \Delta_0 \,\ln\frac{1.14\omega_0}{T} \,g\nu_1(r)
 +\, g T \sum\limits_\omega\, Z_\omega
    \,|G^0_{\omega}(r)|^2  \,.
\eqno(43b)
$$
where
$$
Z_{\omega} = \,{\cal T}_{\omega}\, Y_{\omega}\, +
	     \,{\cal T}_{-\omega}\, Y_{-\omega}\, +
	     \left|{\cal T}_{\omega}\right|^2 \, X_{\omega}\,,
$$
$$
\qquad X_\omega =
\int \, \Delta_1(r) \,\left|G^0_{\omega}(r)\right|^2\, d^3 r
\,, \,
$$
$$
Y_\omega =
\int \, \Delta_1( r)\, G^0_{\omega}(r)\, d^3 r \,.
\eqno(44)
$$
Substituting  $\Delta_1(r)$ from
(43$b$) into expressions
(44), and estimating arising integrals
$$
g\int \nu_1(r) \,|G^0_{\omega}(r)|^2 \,d^3 r \equiv
\lambda_{01}\,,\qquad
$$
$$
g\int |G^0_{\omega}(r)|^2 |G^0_{\omega'}(r)|^2 \,d^3 r \equiv
\lambda_{11}\,,\qquad
$$
$$
g\int \nu_1(r) \,G^0_{\omega}(r) \,d^3 r \equiv
\lambda'_{02} + i \lambda''_{02}\, {\rm sign}\,\omega\,,\qquad
$$
$$
g\int G^0_{\omega}(r) \, |G^0_{\omega'}(r)|^2 \,d^3 r \equiv
\lambda'_{12} + i \lambda''_{12}\, {\rm sign}\,\omega\,
\eqno(45)
$$
with the use of expressions for $G^0_\omega(r)$ and $\nu_1(r)$
(where the real and imaginary parts are denoted by a prime
and two primes)
$$
 G^0_\omega(r) = -\frac{\pi \nu_0}{k_F r}
\exp\left\{-\frac{|\omega|}{v_F} r+ik_F r\,{\rm sign}\,
\omega \right\}
$$
$$
 \nu_1(r) = -\pi \nu_0^2\,\frac{{\cal T}''\cos{2k_F r} +
 {\cal T}'\sin{2k_F r}}{(k_F r)^2}
 \,,\eqno(46)
$$
it is easy to  see
that the integrals converge
already for $\omega,\,\,\omega'=0$,
so parameters  $\lambda_{01}$, $\lambda_{11}$, etc. can
be considered as constant; it allows to write (44)
in the form
$$
X_\omega = X\,, \qquad
Y_\omega= Y' + i Y''\, {\rm sign}\,\omega\,,
$$
$$
Z_{\omega} = 2\,{\cal T}'_{\omega}\, Y'\, -
         2\,|{\cal T}''_{\omega}|\, Y''\, +
         \left|{\cal T}_{\omega}\right|^2 \, X\,.
\eqno(47)
$$

\vspace{3mm}

{\it The region remote from the resonance.} The natural scale for
the energy dependence of  ${\cal T}$-matrix is given by the
bandwidth $J$, so ${\cal T}'_\omega$ and $|{\cal T}''_\omega|$ can
be considered as independent of $\omega$ anywhere, excepting the
vicinity of the resonance (see below). Then $Z_\omega$ is also
independent of  $\omega$, and substitution of $\Delta_1(r)$ from
(43$b$) into  (44) leads to the linear system of equations for
$\Delta_0$ and $Z$ (see Appendix), whose solubility condition
gives
$$
\frac{\delta T_c}{T_{c0}} = \frac{1}{\lambda_0 L^3}\,
\int d^3r \, \frac{ \nu_0  \nu_1(r) +  \nu_1(r)^2}{\nu_0^2}
\,.
\eqno(48)
$$
Eq.\,48 is a natural generalization of the result (1):
the first term in the numerator corresponds to the Anderson
theorem, while the second determines corrections to it.
A configuration of the order parameter shows that (48)
corresponds to the delocalized regime.

For weak impurities ($|V|\ll J$) one has the estimates
$$
{\cal T}'\approx {\cal T} \sim Va^3\,,\qquad
{\cal T}''\sim V a^3 (V/J) (k_F a) \,,
\eqno(49)
$$
and
$$
\nu_0\langle\nu_1\rangle_{0} \sim a^3 \nu_0^2 \,
(V/J)(k_F a)^{-2} \,,
$$
$$
\langle\nu_1^2 \rangle_{0} \sim a^3 \nu_0^2
\,(V^2/J^2)(k_F a)^{-1}   \,,
\eqno(50)
$$
so the Anderson term is leading both in parameter  $V/J$
and in parameter $(k_F a)^{-1}$.
We accepted here $k_Fa\ll 1$, having in mind a situation
near the band end, while estimates for
the band center
follow at  $k_F a\sim 1$.

The delocalized regime retains in the case when  the
resonance condition $1\approx V I(\epsilon_F)$ is formally
fulfilled, but the density of states $\nu(\epsilon_F)$ is
sufficiently large to provide a strong attenuation of the
quasi-local state. In this situation
$$
{\cal T}'\sim J a^3 \frac{J\epsilon_0}{\gamma^2}\,,\qquad
{\cal T}''\sim J a^3 \frac{J}{\gamma} \,,
 \eqno(51)
 $$
and one has under condition  $\gamma\agt \epsilon_0$ (where
$\epsilon_0$ and  $\gamma$ are defined in Eq. 54)
$$
\nu_0\,\langle\nu_1\rangle_{0} \sim
\langle\nu_1^2\rangle_{0} \sim a^3 \nu_0^2\, (k_F a)^{-2} \,,
\eqno(52)
 $$
i.e. the Anderson term has the same order, as a correction to it.

\vspace{3mm}

{\it Vicinity of the resonance.} If $\epsilon_0$ is a root of
equation $1=VI(\epsilon)$, then in the vicinity of it
$$
1-VI(\epsilon)=(\epsilon-\epsilon_0)/E_0\,, \qquad E_0\sim J\,,
\eqno(53)
$$
and hence
$$
{\cal T}=\frac{VE_0}{\epsilon-\epsilon_0+i\gamma}\,,
\qquad \gamma=\pi VE_0 \nu(\epsilon_0) \,.
\eqno(54)
$$
In the Matsubara representation one has
$$
{\cal T}_\omega=\frac{VE_0}{i\omega-\epsilon_0 +i\gamma\,{\rm
sign}\,\omega}\,\equiv
{\cal T}'_\omega - i |{\cal T}''_\omega| {\rm sign}\,\omega\,,
\eqno(55)
$$
where
$$
{\cal T}'_\omega=
-VE_0\,\frac{\epsilon_0}{\epsilon^2_0+(|\omega|+\gamma)^2}\,,
$$
$$
\qquad\qquad
|{\cal T}''_\omega|=
VE_0\,\frac{|\omega|+\gamma}{\epsilon^2_0+(|\omega|+\gamma)^2}\,,
\eqno(56)
$$
so ${\cal T}$-matrix can be considered as independent of
$\omega$ under condition
$$
|\omega| \ll \gamma\,, \qquad {\rm or} \qquad
k_F a \gg \omega/J   \,,
\eqno(57)
$$
i.e. not very close to the initial band edge. If this condition
is not fulfilled\,\footnote{\,In this case, the factor
$\exp(-|\omega|r/v_F)$ restricts contribution to the integrals
(45) by the atomic scale, where expressions (46) are
inapplicable.}, then the $\omega$ dependence   is essential
for the quantities  ${\cal T}'_\omega$  and $|{\cal
T}''_\omega|$, and hence for $Z_\omega$.
In fact, only one combination is relevant,
$$
S=   T \sum\limits_\omega\, Z_\omega \,, \qquad
 \eqno(58)
$$
and substitution of $\Delta_1(r)$ from (43$b$) into
 (44) allows to express  $Z_\omega$ through $\Delta_0$
 and $S$; substituting this expression for $Z_\omega$ into  (58)
 and (43), one comes to the linear system of equations for
 $\Delta_0$ and $S$; its solubility condition with only leading
 terms retained  (see Appendix) reduces to
$$
\left(\tau -
\frac{\nu_0\langle\nu_1\rangle_{0}\! +\!
\langle\nu_1^2\rangle_{0}}{\lambda_0 L^3 \,\nu_0^2}\right)
 \left(-1 + \lambda_{11} V^2 E_0^2 A(T)\right)+
$$
$$
\qquad +\, \tau_c^2 =0\,,
 \eqno(59)
$$
where   $A(T)$ corresponds to expression (24).
Equation  (59) describes the typical situation
related with  intersection of terms. The zero of the first
bracket corresponds to the delocalized regime (see Eq.48),
and vanishing
of  the second bracket corresponds to equation
for $T_c$  of the localized superconductivity
(compare with  (17, 24)),  while the term $\tau_c^2\sim (a/L)^{3}$
removes the degeneracy of terms in the intersection point
(Fig.7).
\begin{figure*}
\centerline{\includegraphics[width=4.8 in]{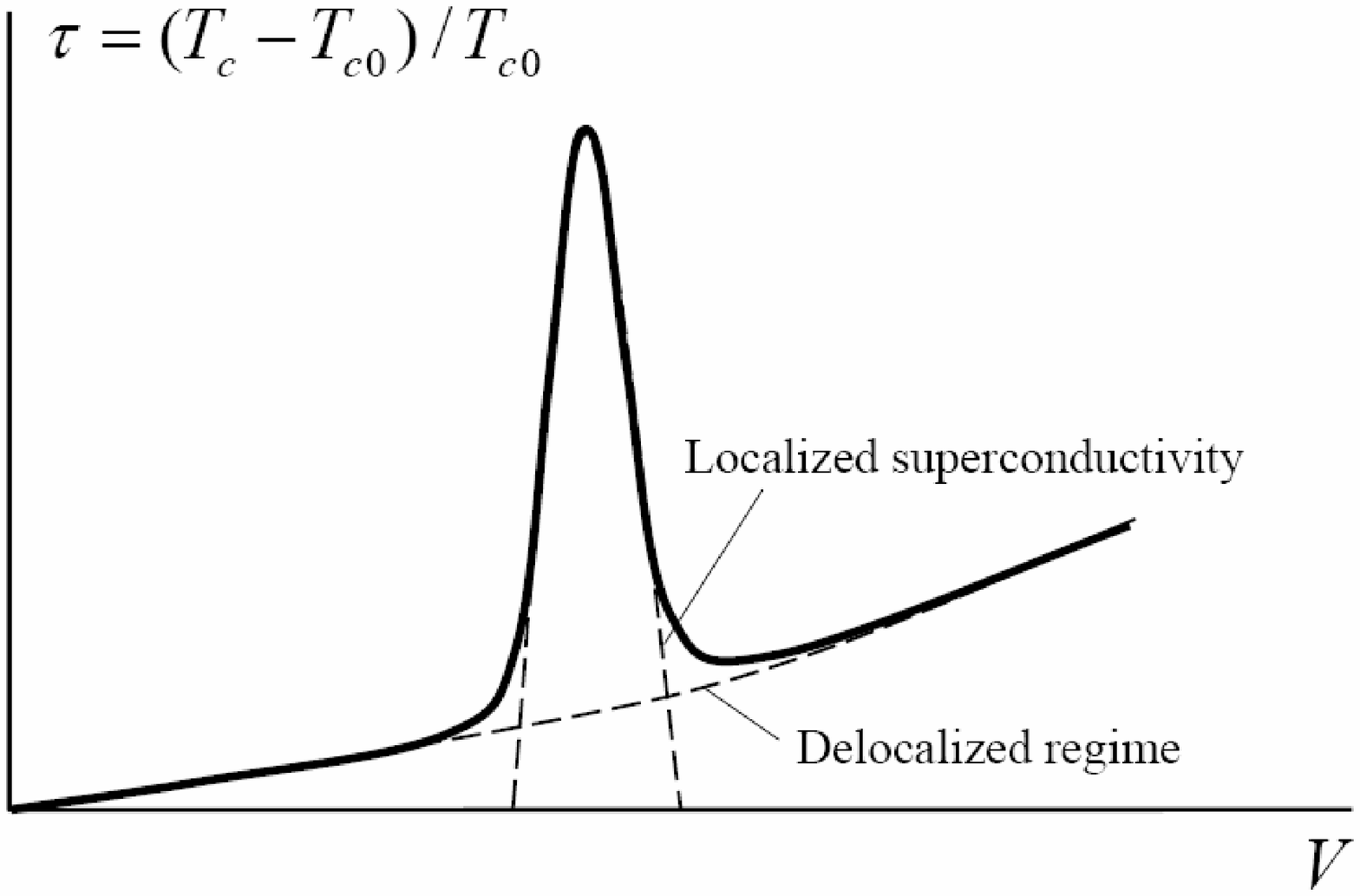}}
\caption{\footnotesize Dependence of  $T_c$ on the impurity
potential  $V$ for a small impurity concentration. } \label{fig7}
\end{figure*}

\begin{center}
{\bf 5. Consequences for the Anderson model.} \end{center}

Usually localization is studied in the framework of the Anderson
model, which is a discrete version of the Schroedinger equation
with a random potential: the bare spectrum is a band of width $J$,
while the potential values $V_n$ on the lattice sites are
independent random quantities with the distribution  $P\{V\}$ of
width  $\sim W$, which is supposed to be rectangular. To transfer
from the one-impurity problem to the Anderson model, it is
sufficient to accept that a potential  $V$ of impurities
fluctuates in the interval $(-W,W)$, while their concentration
$c$ is gradually increased from small values to  values of the
order of unity.

The results of Sec.4 correspond formally to the periodic
arrangement of impurities, but in fact their periodicity is not
essential:  each impurity arouses only a local deformation of the
order parameter and these deformations are independent in case
of a small concentration. If $\Delta_1(r)$  and $\nu_1(r)$
correspond to the one-impurity problem, then configurations
$$
\Delta(r)=\Delta_0+ \sum\limits_i \Delta_1(r-r_i)\,,
$$
$$
\nu_F(r)=\nu_0+ \sum\limits_i \nu_1(r-r_i)\,
\eqno(60)
$$
correspond to a situation, when  several impurities  are
arranged in points $r_i$:  it is a consequence of localization
of the kernel $K_1(r,r')$ in both variables near the defect
position. It is clear from (41) that the amplitude of
$\Delta_1(r)$ is proportional to  $\Delta_0$, so
$\Delta_1(r)=\Delta_0 f(r)$ and substitution of (60)
into  (10) gives for  $T_c$ close to  $T_{c0}$:
$$
\frac{\delta T_c}{T_{c0}} = \frac{m}{\lambda_0^2 L^3}\, g \!
\int d^3r \left[ \nu_1(r) + \nu_1(r) f(r) \right]\,.
\eqno(61)
$$
Here  $m$ is a number of impurities in the volume $L^3$,
and $f(r)$ can be identified as $\nu_1(r)/\nu_0$ from comparison
with (48)\,\footnote{\,For weak disorder, relation
$\Delta_1(r)=\Delta_0 \nu_1(r)/\nu_0$ follows from the second
equation (41) after neglecting the quantity  $F(r)$,
which is of the second order. Its validity for the
delocalized regime without assumption of small $\nu_1(r)$
is a non-trivial result expressed by equation (48).  }. One can
see that effect is proportional to a concentration of impurities,
while their arrangement is irrelevant. The Anderson theorem
is valid under  condition $|f(r)|\ll 1$, which
is fulfilled for weak impurities. In  case of nonequivalent
impurities, the result  (61) should be averaged according to
distribution  $P(V)$.

In the regime of the  localized order parameter, each impurity is
practically independent of enviroment and $T_c$ of the system is
determined by those of them, which are close to a
resonance; if distribution $P(V)$ is continuous and sufficiently
wide, then the condition of almost  exact resonance is always
realized with a certain probability. Therefore, the concentration
of the resonant impurities is finite and their quasiperiodic
arrangement stabilizes the mean-field solution.

Above considerations completely clarify a situation for small
impurity concentrations. Advancement to higher concentrations is
simplified by observation that equations  (31), (32) never used a
fact that  $G^0_{nn'}$ corresponds to an ideal lattice; the same
equations describe insertion of an additional impurity in the
disordered superconductor. Noticing that
$$
G^0_{n_0 n_0}= \sum\limits_s  \frac{|\varphi_s(n_0)|^2}
{E-\epsilon_s+i 0}
=\int\limits d\epsilon  \frac{\nu(\epsilon, n_0)}
{E-\epsilon+i 0}
$$
and replacing  $\nu(\epsilon, n_0)$ by its mean value
$\langle \nu(\epsilon)\rangle$, we obtain the same
representation $I(E) -i \pi \nu(E)$ as in Eq.33, with
a predictable behavior of   $I(E)$ and $\nu(E)$ (Fig.5,$b$).
\vspace{3mm}

{\it Weak impurities.} In this case, a behavior of functions
$I(E)$ and  $\nu(E)$ differs from their behavior in an ideal
crystal by small smoothening of the Van Hove singularities
(Fig.5,$b$). Dependence on $n_0$ results in
fluctuations of the
form of these functions, which are also small. It is clear that
for weak impurities ($|V|\ll J$) the resonance condition  is not
fulfilled and no localization of the order parameter is possible.

For the delocalized regime, it is convenient to present the
result (61) in another form. Taking
the one-impurity configuration
$\Delta(r)= \Delta_0+\Delta_0 f(r)$, $\nu(r)= \nu_0+\nu_1(r)$
and substituting it into equation (10), we have for the effective
density of states entering into the BCS formula:
$$
\nu_{eff}=
\frac{\nu_0 +\nu_0 \langle f \rangle + \langle \nu_1
  \rangle + \langle \nu_1 f \rangle
}{1+\langle f \rangle} \,.
\eqno(62)
$$
Subtracting the result with $f\equiv 0$ and retaining
the main terms   in $L^{-d}$:
$$
\nu_{eff}- \langle \nu \rangle \,= \, \langle \nu_1 f \rangle
\,= \,\frac{ \langle \nu^2_1 f \rangle}{\nu_0} \,= \,
 \frac{\,\,\nu_0 \left( {\cal T}'\right)^2}{4k_F L^3} \,,
$$
$$
\langle \nu \rangle -\nu_0\, =\,  \langle \nu_1  \rangle
\,=\, - \,  \frac{(\nu_0{\cal T}')}{2k_F^2 L^3}   \,,
\eqno(63)
$$
where we have taken into account that only the term with
${\cal T}' \approx V+V^2I(E_F)$ is essential in Eq.46
for weak impurities.  Inserting impurities one after
another and averaging over  $V$,
$$
\nu_{eff}- \langle \nu \rangle  \sim
c \nu_0 \, \frac{W^2}{J^2} \, (k_F a)^{-1}
\,,
$$
$$
\langle \nu \rangle -\nu_0  \sim
c \nu_0 \, \frac{W^2}{J^2} \, (k_F a)^{-2} \,,  \,\,
\eqno(64)
$$
we see that in the course of increasing a
concentration, the increment of the quantity $\nu_{eff}- \langle
\nu \rangle$  is by a factor $k_Fa$ smaller than the increment of
$\langle \nu \rangle -\nu_0$. Near the band edge one have
$\langle \nu \rangle \gg \nu_0$ for sufficiently large
concentrations, so
$\nu_{eff}-\langle \nu \rangle \sim k_Fa \langle \nu \rangle$
and deviations from the Anderson theorem are small. Near the band
center we have $k_Fa \sim 1$ and differencies
$\langle \nu \rangle - \nu_{0}$ and $\nu_{eff}-\langle \nu \rangle$
are small till concentrations $c\sim 1$; so
$\nu_{eff}- \langle \nu \rangle \ll \langle \nu \rangle$.
It is clear that violation of  self-averaging
for the order parameter does not occur for weak impurities.

In the $3D$ case, isolated weak impurities do not produce bound
states  beyond the initial spectrum  (it is clear from Fig.5,$a$),
and a finite density of states in this energy interval is a
collective phenomenon related with long-range fluctuations of the
band edge. Consider a fluctuation in the region of size $L$, due
to which the range of $V$ values is somewhat restricted, $(-W,
W-2\delta)$ instead $(-W,W)$. Then the mean value of the random
potential is decreased by a quantity  $\delta$, while a
probability of such fluctuation

\noindent
$\exp(-L^d \delta/W)$ is not
small for  $L^d \delta/W\alt 1$. Such fluctuations occur at all
scales and produce a finite density of states beyond the bare
spectrum.\,\footnote{\,The amplitude of long-range fluctuations
can be seen from the fact that in the extremal cases the whole
band is shifted by a quantity $W$ or $-W$, i.e. such fluctuations by
themselves (with no account for partial discretization of
spectrum) cannot produce unbounded values of $\nu_F(r)$. } If size
$L$ of the fluctuation is sufficient for existing of
superconductivity, the latter will not differ from
superconductivity in the initial system with not shifted band
edge; i.e. impurities will not violate the uniformity of the order
parameter. Near the initial band edge, the indicated fluctuations
strongly overlap and superconductivity is quasi-homogeneous. Such
fluctuations become spatially isolated in the region of strong
localization, where they can be described in terms of the size
effect (Sec.6).

\vspace{3mm}
{\it Strong impurities.} For a small concentration of strong
impurities  ($|V| \agt J$), a behavior of functions  $I(E)$ and
$\nu(E)$ is not very different from their
behavior in an ideal crystal. However, in the region of the
maximum of $I(E)$ the density of states becomes finite and,
at first glance,  complicates the occurrence of resonances.
In fact, a new phenomenon comes to life.
Since now  $G^0_{n_0 n_0}$ depends on  $n_0$, the attenuation
of the quasi-local state will be determined not by average
density of states, but its local value at the point $n_0$,
which can be small in a fluctuational manner. As a result,
resonances become possible even for energies
in the deep of  the band,
where they were forbidden in the ideal lattice.
The typical situation, when the local density of states
$\nu_F(n_0)$ is small, corresponds to large values of the
random potential in the vicinity of $n_0$; if now
an impurity with large negative $V$ is inserted into the site
$n_0$, then a specific
%
\begin{figure}
\centerline{\includegraphics[width=2.7 in]{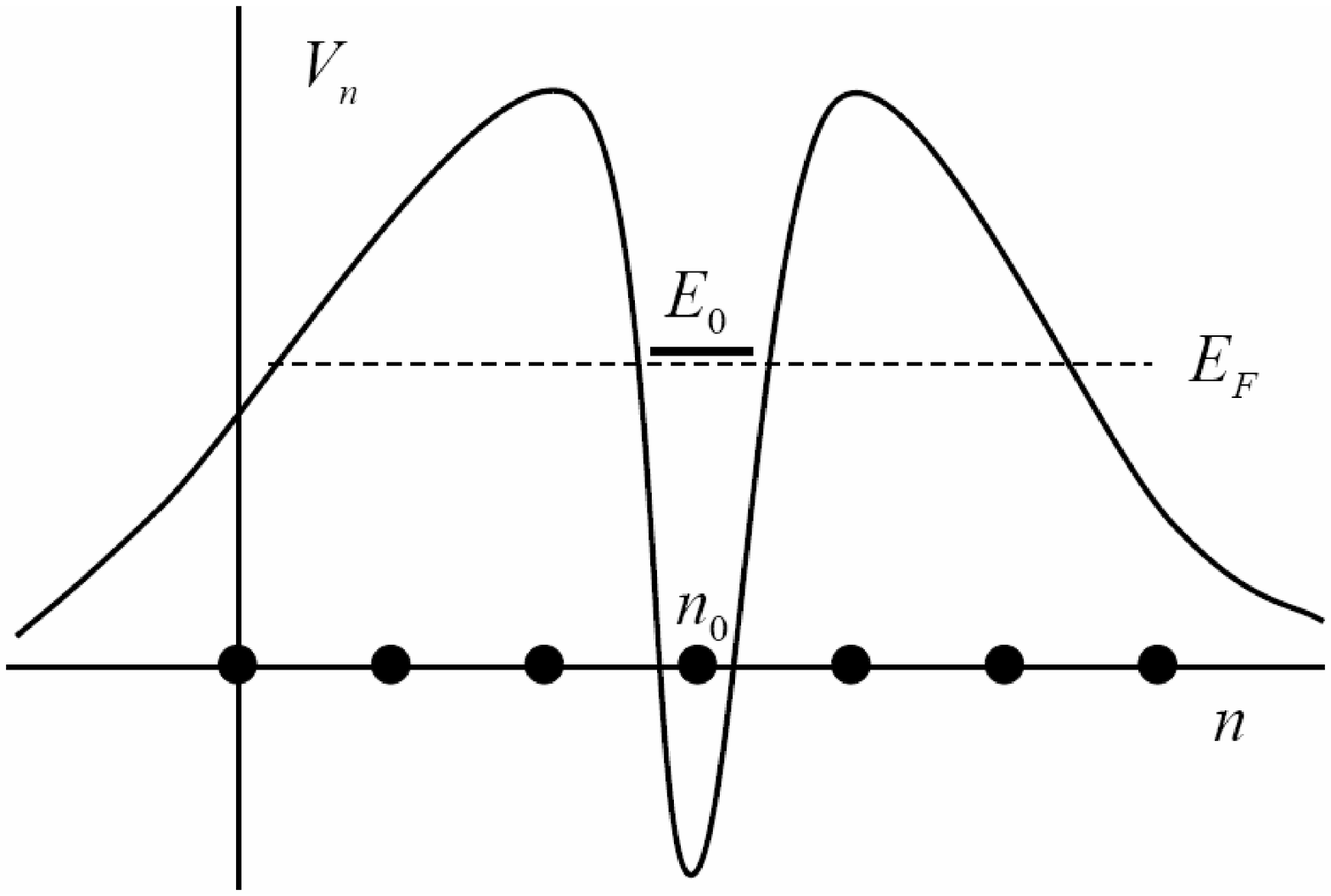}}
\caption{\footnotesize A typical fluctuation of the random
potential responsible for  existence of the quasi-local state
in the deep of the allowed band.} \label{fig8}
\end{figure}
resonance configuration arises (Fig.8).\,\footnote{\,According to
\cite{304}, such configurations are responsible for
multifractal statistics. It appears, that the tails of the
distribution function are determined by individual peaks (and not
fractal clusters), in correspondence with our conception.
  Thereby, we do not ignore the existence of
multifractality but give another description of its influence on
superconductivity. } In the "minimal" variant, such configuration
corresponds to existence of large barriers at the nearest
neighbours of  site  $n_0$, while a value of the potential at
$n_0$ is chosen so that a corresponding level was in the
interval of width  $\sim T_c$ near the Fermi energy (the
probability of this event is $\sim T_c/W$).  For a finite band,
both large positive and large negative value of the potential are
locking, and for $W\gg J$ such values occur with probability $p$
close to unity. Therefore, the probability of the "minimal"
fluctuation
$$
P_{res} \sim p^Z \frac{T_c}{W} \,,
\eqno(65)
$$
where  $Z$ is a number of the nearest neighbours.
 It is clear that such resonances can occur for any position
 of the Fermi level. In the  region of the fluctuational tail,
the density of states is small by the natural reasons and there is
no need to create the barrier around $n_0$; so the factor  $p^Z$
will be absent but the less probable form of the effective
potential well is necessary, in order the level was in the
desired part of the spectrum.\,\footnote{\,Strictly
speaking, the resonant configurations of such kind are
possible for small $W$ in the vicinity of the initial band
edge.  However, a size of such configurations is inevitably
large (due to restriction of the barrier height and
absence of levels in a shallow well of a small radius), so
they have a negligible probability and are incompatible
with electroneutrality (Sec.3).  } With increasing of the
impurity concentration, the effective bandwidth is extended and
the maximum of $I(E)$  is shifted correspondingly. However,
the general mechanism of resonances and estimation of their
probability remain unchanged.

Since the true critical temperature is hardly observable, it is
actual to consider the "bulk $T_c$", which can be defined as $T_c$
of the system with excluded resonant impurities. For strong but
not resonant impurities, two terms in Eq.48 are of the same order
(see (50,\,52)), and impurities are independent till
concentrations  $c\sim 1$, since the mobility edge lies far from
the bare edge of spectrum and  $k_F a\sim 1$. Validity of the
Anderson theorem holds on the qualitative level:  $T_c$ is
determined by the effective density of states, which differs from
the average one by a factor of the order of unity.

\begin{center}
{\bf 6. Size effect in the localized phase.}
\end{center}

In the localized phase, the system breaks up into
quasi-independent blocks of size $\xi$, and superconductivity is
suppressed due to the size effect.
Below we analyze this effect in terms of the
Gor'kov equation. Superconductivity in small samples was discussed
in many papers  (see a review article \cite{26}), but this
discussion mainly concerns the aspects:

(a) inadequacy of the grand canonical ensemble due
to a fixed number of electrons in small granules;

(b) parity effects;

(c) insufficiency of the mean field approximation;

(d) absence of an abrupt phase transition, etc.

\noindent
which
 are essential for finite
systems and completely not actual in the present context.
In principle, it is  correct to stress unreliebility of the mean
field approach, but all attempts to overcome it
%
%
(from modified mean field approximations till the exact
Richardson solution and a direct numerical modelling) are
based on the truncated BCS Hamiltonian, which by itself
induces the certain way
  of pairing (in general incorrect)\,\footnote{\,The
state $\varphi_s$ is coupled
with its complex conjugated: it is correct only
for a uniform order parameter \cite{15}.  }.
As for the Gor'kov equation, it
corresponds to the saddle-point approximation in the
functional integral \cite{27a,27} and is the most
grounded
of all mean-field type approaches; in addition,
the electron interaction is specified
in the physically clear manner and
independently of one-electron states  (Sec.2). The accuracy of
approximation is determined by the Ginzburg parameter, which
provides insignificance of fluctuations in  case of a
 superconductor (with exception of some special cases:
e.g. in finite systems fluctuations have a qualitative
importance, destroying a phase transition). The Gor'kov
equation can be also obtained from the Eliashberg equations
in the limit of the local interaction \cite{14}.

Consider the cubic sample of size  $L$, accepting the
periodical boundary conditions for the electron
eigenfunctions. In a pure superconductor the latter have a form
of  plain waves, so $|\varphi_s(r)|^2=L^{-d}$  and the local
density of states  (8) does not depend on  $r$. Then
$\Delta(r)=const$  is an exact solution of the Gor'kov equation
(4), which reduces to
$$
\Delta = gT \sum \limits_{\omega} L^{-d}\sum\limits_s
\frac{1}{\epsilon_s^2+\omega^2} \Delta
\eqno(66)
 $$
and coincides with (9) in case of the continuous spectrum.
In a small energy interval, the spectrum can be considered
as a set of equidistant levels with a spacing  $\Omega$
$$
\epsilon_s=\Omega (s+1/2)\,, \qquad \Omega=1/\nu_F
L^d \,,\eqno(67)
$$
where we accept that the Fermi energy lies in the middle
between two discrete levels\,\footnote{\,Such assumption is
commonly accepted  \cite{26} for the case of the even number
of electrons  $N$; for odd $N$ it is accepted
$\epsilon_s=\Omega s$, but the level $\epsilon_0=0$
is considered as "blocked", i.e. occupied by the unpaired
electron and not participating in the scattering process.
In the latter case, the results are analogous but correspond
to smaller $T_c$.  }.
Substitution to (66) and summation over $s$ gives
$$
1=g T \sum \limits_{\omega} \frac{\pi \nu_F }{|\omega|}
\tanh\frac{\pi| \omega|}{\Omega}   \,.
 \eqno(68)
 $$
For small $\Omega$, the argument of the hyperbolic tangent
is large and one can set $\tanh{x}= 1-2 e^{-2x}$, so
$$
\frac{1}{g\nu_F}=  \ln\frac{1.14\omega_0}{T}
-4e^{-2\pi^2T/\Omega} \,,
 \eqno(69)
 $$
where we retained only  main terms with $\omega=\pm \pi T$
in the second sum over  $\omega$. Subtracting the analogous
equation with  $\Omega=0$, it is easy to obtain
$$
T_c= T_{c0} \left[ 1 - 4e^{-2\pi^2T_{c0}/\Omega} \right]\,,
\qquad   \Omega\ll T_{c0}  \,.
 \eqno(70)
 $$
For  $T\to 0$, one can replace summation in (68)  by
integration and obtain the equation for the critical value
of $\Omega$, at which  superconductivity is destroyed
$$
\frac{1}{g\nu_F}
=  \int_{-\omega_0}^{\omega_0} \frac{ d\omega
}{2\omega} \tanh\frac{\pi  \omega}{\Omega_c} =
$$
$$=
 \ln \frac{\pi  \omega_0}{\Omega_c} - \int_{0}^\infty
 \frac{\ln x}{\cosh^2 x} dx  \,.
\eqno(71)
$$
The last integral is equal $\ln(\pi/4\gamma)$, where
$\ln\gamma=C=0.577$  is the Euler constant and comparing with
the result for  $T_{c0}$
$$
1 =g\nu_F \ln\frac{4\gamma \omega_0}{\Omega_c} \,,\qquad
1 =g\nu_F \ln\frac{2\gamma \omega_0}{\pi T_{c0}} \,,
 \eqno(72)
$$
one can see that
$$
\Omega_c= 2 \pi  T_{c0} \,.
 \eqno(73)
$$
To find the dependence of  $T_c$ on $\Omega$ in the vicinity of
$\Omega_c$, one can transfer (68) using the Poisson summation
formula \cite{23}
$$
\frac{1}{g\nu_F} = \sum\limits_{s=-\infty}^\infty
 e^{-i\pi s}
 \int\limits_{-\pi\omega_0/\Omega}^{\pi\omega_0/\Omega}
 d x \,\frac{  \tanh x}{2x }
\exp\left\{i\frac{ s\Omega }{\pi T} x \right\}   \,,
\eqno(74)
$$
where the term with $s=0$ corresponds to (71). For
$s\ne 0$, the integrals are convergent at large $|x|$ and
it is possible to set
$\omega_0=\infty$ in them. Due to evenness in
$s$ they can be calculated for $s>0$; then the contour is shifted
in the upper half-plain and the main contribution arises from the
pole  $x=i\pi/2$.  For $\Omega/T\gg 1$ it is  sufficient to
retain the terms with  $s=0,\pm 1$,
$$
\frac{1}{g\nu_F}=  \ln\frac{4\gamma\omega_0}{\Omega}
-4e^{-\Omega/2T} \,,
 \eqno(75)
 $$
and subtracting the analogous equation with $T=0$
$$
T_c=\frac{\Omega_c}{2\ln\left[4 \Omega_c/(\Omega_c-\Omega) \right]
 } \,, \qquad  \Omega\to \Omega_c
 \eqno(76)
 $$
In the reduced coordinates
$$
y=T_c/T_{c0}\,,  \qquad x= \Omega/\Omega_c
\eqno(77)
$$
one can obtain the universal dependence $y(x)$.
Indeed, transforming (68) by subtraction of the analogous
equation with  $\Omega=0$,  one has
$$
\ln\frac{T}{T_{c0}} = T \sum \limits_{\omega} \frac{\pi
}{|\omega|} \left(\tanh\frac{\pi| \omega|}{\Omega} -1
\right)  \,,
\eqno(78)
$$
where  $\omega_0$ can be set to infinity. Substituting
the Matsubara  values $\pi T (2n+1)$  for $\omega$, one can
present the dependence $y(x)$ in the parametric form
$$
y=\exp F(t)\,,\qquad x= t^{-1} \exp F(t)\,,
$$
$$
F(t) = 2 \sum \limits_{n=0}^\infty \frac{1}{2n+1}
\left[\tanh\frac{\pi (2n+1) t}{2} -1 \right]  \,,
\eqno(79)
$$
where $t$ runs from zero to infinity. Numerical calculation
based on  (79) gives the "rectangular" dependence
$y(x)$ shown in Fig.9:  this dependence has exponentially small
deviation from the horizontal  line near $y=1$, and
exponentially small deviation from the vertical  line
near  $x=1$.
\begin{figure}
\centerline{\includegraphics[width=2.5 in]{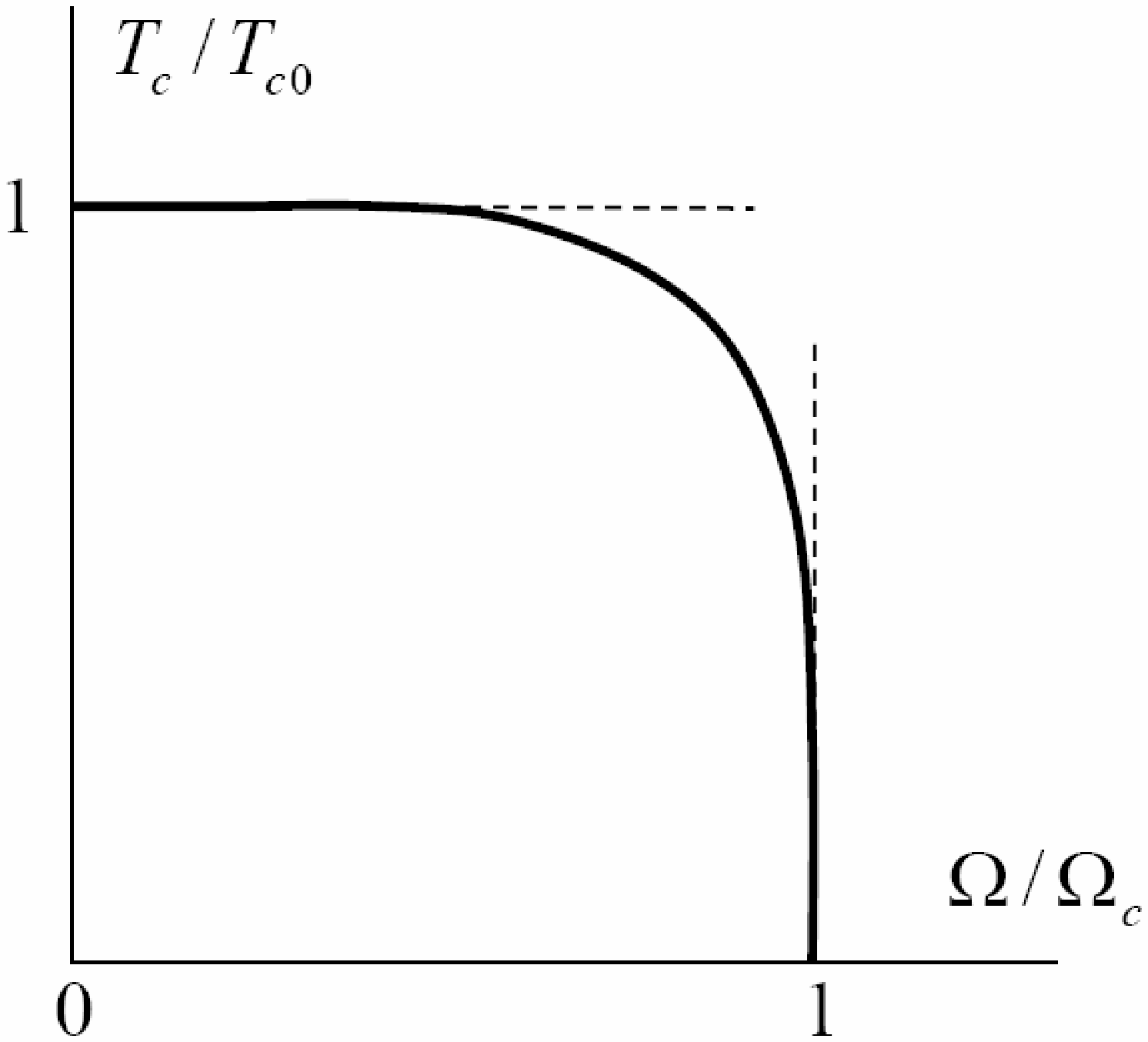}}
\caption{\footnotesize "Rectangular" dependence of  $T_c$
on the level spacing  $\Omega$ in a finite system; it is
universal in the reduced coordinates  $ y=T_c/T_{c0}$,
$ x= \Omega/\Omega_c$. } \label{fig9}
\end{figure}

The given consideration retains for a disordered superconductor
if possibility of self-averaging is accepted.\,\footnote{\,Of
course, in this case one should take some realistic statistics of
the Wigner--Dyson kind instead of the equidistant levels, but
it has a small effect on the results \cite{26}.  }.
The obtained results can be used to describe the
dependence  of $T_c$ on  the distance to the
mobility edge in the localized phase, where the system is
divided into quasi-independent blocks of size  $\xi$. The role of
$\Omega$ is played by the quantity
$$
\Omega(E) \sim J (\xi/a)^{-d} \sim J (|E-E_c|/J)^{d\nu} \,,
\eqno(80)
$$
where  $\nu$ is the critical exponent of the localization length.
According to Sec.5, the assumption of self-averaging is valid
literally for weak disorder and on the qualitative level for
strong disorder in the absence of  resonances. In the latter
case, $T_c$ is determined by the effective density of states
which differs from the average one by a factor of the order of
unity, which is a smooth function of parameters.  It preserves
the character of singularities  (70) and (76), which determine
the behavior near  $E_c$ and $E^*$ (Fig.3) and are responsible
for the most striking features in the dependence
$T_c(\epsilon_F)$.

\begin{center}
{\bf 7. Conclusion.}
\end{center}

The present paper resolves contradiction between two
series of papers \cite{1}--\cite{5} and \cite{9,10}. The
obtained results has in some way a compromise character.
On the one hand, the "bulk" superconductivity behaves in
correspondence with the picture by Bulaevskii and Sadovskii
\cite{1}--\cite{5}.  On the other hand, the true transition
temperature $T_c$ of strongly disordered superconductor
does not coincide with the "bulk" one and is determined by
rare peaks of the order parameter on the atomic scale; in
correspondence with  \cite{9,10} it has a power law dependence
on the coupling constant and does not depend on the cut-off
frequency.  However, in contrast to \cite{9,10}, it has
no essential dependence on the position of the Fermi
level and does not correlate with the Anderson transition.
By this reason, we do not see any grounds to say on
"fractal superconductivity" \cite{10} near the localization
threshold.

The obtained results are obtained in the framework of the
mean field theory, which is surely valid in the delocalized
regime. In fluctuational theory, essential modification
of results  is expected only for the localized
regime: the modulus of the order parameter changes slightly,
while fluctuations of its phase become essential.
We should stress  that the role of
fluctuations is determined by specific values of
parameters, characterizing the system:
 if, for example, the ratio
$T_c/J$ is not too small, then the resonant impurities have
rather large concentration and the Josephson coupling between
the localized superconducting "drops" is sufficiently large for
stabilization of the mean-field solution (this coupling is
determined mainly by existence of the uniform contribution
(see (30)), which grows at small $L$).
Contrary, if  $T_c/J\to 0$, then the Josephson coupling
between  drops is small and fluctuations essentially suppress
$T_c$.
According to the nonlinear Ginzburg--Landau equations derived
in  \cite{11}  for the localized regime, decreasing of the
temperature stimulates the growing of tails of the localized
solutions; the Josephson coupling between  drops becomes
greater and stabilizes the mean-field solution before the
"bulk $T_c$" is reached. Analogous remarks
are valid in relation with the Coulomb blocade effects \cite{27a}.

In comparison of the obtained results with experiment, one
should have in mind, that the continuous distribution $P(V)$ in the
Anderson model is not very realistic; it is more adequate to assume
the discrete (and not very dense) set of the  $V$ values.
As a result, in  most  systems the described resonanses
will be unobservable for any concentration and arrangement of
impurities. However, in the minority of systems the
effect of resonances will be strong and stable. The Anderson
model with a several types of periodically arranged impurities
can be considered as the model for the high-temperature oxide
superconductors. The possibility to interpret the  "superconducting
explosion" of 1987 as localization of the order parameter
was indicated previously  \cite{11}; the above results suggests
possibility of such localization not only in the $Cu\!-\!O$ planes
but also at the individual atoms. The adequacy of such a model
is confirmed by (a) optimistic estimates of $T_c$, (b) practical
coincidence of the maximal $T_c$ values  with $\omega_0/\pi$,
(c) suppressed isotop-effect in the regime
$\epsilon_c\ll \omega_0$.

\begin{center}
{\it Appedix. On solution of the Gor'kov equation with the
kernel (37).} \end{center}

Let fill in the gaps for our exposition in the main text.
\vspace{1mm}

In the region remote from the resonance, we can consider ${\cal
T}'_\omega$  and $|{\cal T}''_\omega|$ as independent of
$\omega$: then  $Z_\omega$ is also constant. Substituting
$\Delta_1(r)$ from (43$b$) into expressions  (44) for $X_\omega$
and $Y_\omega$, we have representation  (47) with parameters
$$
X= \Delta_0 \ln\frac{1.14\omega_0}{T} \,\lambda_{01}
   + \frac{\omega_0}{\pi} \lambda_{11} Z \qquad \, \,
$$
$$
Y'= \Delta_0 \ln\frac{1.14\omega_0}{T} \,\lambda'_{02}
   + \frac{\omega_0}{\pi} \lambda'_{12} Z  \qquad \,\,
$$
$$
Y''= \Delta_0 \ln\frac{1.14\omega_0}{T} \,\lambda''_{02}
+ \frac{\omega_0}{\pi} \lambda''_{12} Z    \,.
\eqno(A.1)
$$
Then  $Z_\omega$ has a form
$$
Z= \Delta_0 \ln\frac{1.14\omega_0}{T} \, B_2   +
Z \frac{\omega_0}{\pi} \,B_3 \,,
\eqno(A.2)
$$
and its combination with (43$a$)  gives a system of equations
for  $\Delta_0$ and $Z$
$$
\Delta_0\left[\,-\frac{1}{B_1} + g \langle\nu_1\rangle_{0}
+g^2 \langle\nu_1^2\rangle_{0} \ln\frac{1.14\omega_0}{T}
\,\right] +
$$
$$
+g\frac{\omega_0}{\pi}  \lambda_{01} Z = 0
$$
$$
\Delta_0 B_2 \ln\frac{1.14\omega_0}{T} +Z\left[\,-1 +
\frac{\omega_0}{\pi} \,B_3\right]
 = 0  \,,
 \eqno(A.3)
$$
with the coefficients
$$
B_1= \frac{K_0(0)}{\lambda_0 L^3 \tau} \ln\frac{1.14\omega_0}{T}
\qquad\qquad\qquad   \,\,  \,\, \,\,
$$
$$
B_2=  2{\cal T}'\lambda'_{02} - 2{\cal T}''\lambda''_{02}
 + |{\cal T}|^2 \lambda_{01}  \qquad\quad
$$
$$
B_3= 2{\cal T}'\lambda'_{12}
   - 2|{\cal T}''|\lambda''_{12}
    + |{\cal T}|^2 \lambda_{11}
 \,.
\eqno(A.4)
$$
The terms containing
$\omega_0$ has an additional smallness   $\sim  \omega_0/J$ and
can be neglected\,\footnote{\,We have in mind the traditional
superconductors. If $\omega_0\sim J$, then the
"vicinity of the resonance" is extended and in fact
occupies
the whole band.}; the condition of solubility for  ($A.3$)
gives the result  (48).

\vspace{3mm}

In the vicinity of the resonance, one cannot neglect
the $\omega$  dependence of the quantities  ${\cal
T}'_\omega$,  $|{\cal T}''_\omega|$, and consequently  $Z_\omega$.
Substituting  $\Delta_1(r)$ from (43$b$) into  (44) for
$X_\omega$  and $Y_\omega$, one has  representation  (47)
with parameters
$$
X= \Delta_0 \ln\frac{1.14\omega_0}{T} \,\lambda_{01}
   +  \lambda_{11} S \qquad \,\,
$$
$$
Y'= \Delta_0 \ln\frac{1.14\omega_0}{T} \,\lambda'_{02}
   +  \lambda'_{12} S   \,\,\qquad
$$
$$
Y''= \Delta_0 \ln\frac{1.14\omega_0}{T} \,\lambda''_{02}
+  \lambda''_{12} S
\eqno(A.5)
$$
and for  $Z_\omega$
$$
Z_\omega=
\Delta_0 \ln\frac{1.14\omega_0}{T} \,
 \left[ 2{\cal T}'_\omega\lambda'_{02} -
 2|{\cal T}''_\omega|\lambda''_{02}
 \right. +
$$
$$  \left.
 + |{\cal T}_\omega|^2 \lambda_{01}\right]
   + S    \left[ 2{\cal T}'_\omega\lambda'_{12}
- 2|{\cal T}''_\omega|\lambda''_{12}
   +    |{\cal T}_\omega|^2
   \lambda_{11}\right]
   \eqno(A.6)
$$
Substitution into expressions (58)  and  (43) gives a system
of equations for  $\Delta_0$ and $S$
$$
\Delta_0\left[\,-\frac{1}{B_1} + g \langle\nu_1\rangle_{0}
+g^2 \langle\nu_1^2\rangle_{0} \ln\frac{1.14\omega_0}{T}
\,\right] +
$$
$$
\qquad \qquad + g \lambda_{01} S = 0 \,,
$$
$$
\Delta_0  \,
 C_1\ln\frac{1.14\omega_0}{T}+ S\,\left[\,-1 + C_2\right]  = 0
 \,, \eqno(A.7)
 $$
with definitions
$$
C_1=  2\lambda'_{02} \sigma_1- 2\lambda''_{02}\sigma_2
 + \lambda_{01} \sigma_3  \,,\qquad
 $$
 $$
C_2=  2\lambda'_{12} \sigma_1- 2\lambda''_{12}\sigma_2
 + \lambda_{11} \sigma_3  \,,\qquad
$$
$$
\sigma_1=   T \sum\limits_\omega\, {\cal T}'_\omega \,, \quad
\sigma_2=   T \sum\limits_\omega\, |{\cal T}''_\omega| \,, \quad
$$
$$ \qquad
\sigma_3=   T \sum\limits_\omega\, |{\cal T}_\omega|^2 \,.
 \eqno(A.8)
$$
The condition of solubility for the system ($A.7$) gives
$$
\left(\,\frac{1}{B_1} - g \langle\nu_1\rangle_{0}
-g^2 \langle\nu_1^2\rangle_{0} \ln\frac{1.14\omega_0}{T}
\,\right) \,\left(\,-1 + C_2\right) +
$$
$$+
 \,C_1 g \lambda_{01}\ln\frac{1.14\omega_0}{T} = 0 \,
 \eqno(A.9)
$$
Estimations for  $\epsilon_0 \sim \gamma\sim T$ give
$$
\sigma_1 \sim VE_0 \,, \quad
\sigma_2 \sim  VE_0 \ln\frac{\omega_0}{T}  \,, \quad
\sigma_3 \sim \frac{(VE_0)^2}{T}  \,,
 \eqno(A.10)
$$
and allow to retain only the leading  terms in $J/T$; as a result,
Eq.$A.9$ can be written as
$$
\left(\,\tau -\frac{g^2 \nu_0\langle\nu_1\rangle_{0}
+g^2 \langle\nu_1^2\rangle_{0}}{\lambda_0^3
L^3}  \,\right) \,\left(\,-1 +
\lambda_{11}\sigma_3\right) +
$$
$$+
\frac{\lambda_{01}^2}{\lambda_0^3 L^3}\, g\sigma_3 = 0
 \eqno(A.11)
$$
and reduces to a form (59); the last term is essential
only near the intersection point of dashed lines in Fig.7,
when $\epsilon_0 \sim \gamma \sim \epsilon_c$  and  $T$
should be replaced by $\epsilon_c$ in the estimates ($A.10$).

\end{document}